\documentclass[11pt,a4paper]{article}
\usepackage[UKenglish,english]{babel}

\usepackage{jheppub}
\usepackage{lmodern}

\usepackage[]{fontenc}
\usepackage[latin9]{inputenc}
\usepackage{array}
\usepackage{float}
\usepackage{esint}

\makeatletter

\pdfpageheight\paperheight
\pdfpagewidth\paperwidth

\providecommand{\tabularnewline}{\\}

\@ifundefined{textcolor}{}
{%
 \definecolor{BLACK}{gray}{0}
 \definecolor{WHITE}{gray}{1}
 \definecolor{RED}{rgb}{1,0,0}
 \definecolor{GREEN}{rgb}{0,1,0}
 \definecolor{BLUE}{rgb}{0,0,1}
 \definecolor{CYAN}{cmyk}{1,0,0,0}
 \definecolor{MAGENTA}{cmyk}{0,1,0,0}
 \definecolor{YELLOW}{cmyk}{0,0,1,0}
}

\renewcommand\[{\begin{equation}}
\renewcommand\]{\end{equation}} 

\AtBeginDocument{
  
}

\makeatother

\begin{document}
\selectlanguage{UKenglish}

\subheader{CERN-PH-TH/2010-210}
\title{The Imperfect Fluid behind Kinetic~Gravity~Braiding }

\author[a]{Oriol Pujol\`as,}
\emailAdd{pujolas@ifae.es}
\author[b]{Ignacy Sawicki}
\emailAdd{ignacy.sawicki@uni-heidelberg.de}
\author[c]{and Alexander Vikman}
\emailAdd{alexander.vikman@cern.ch}

\affiliation[a]{Departament de F\'isica and IFAE, Universitat Aut\`onoma
de Barcelona, \\08193 Bellaterra, Barcelona, Spain}
\affiliation[b]{Institut f\"ur Theoretische Physik, Ruprecht-Karls-Universit\"at
Heidelberg, Philosophenweg 16, \\69120 Heidelberg, Germany}
\affiliation[c]{Theory Division, CERN, CH-1211 Geneva 23, Switzerland}

\date{\today}
\abstract{
We present a standard hydrodynamical description for non-canonical
scalar field theories with \emph{kinetic gravity braiding}. In particular,
this picture applies to the simplest \emph{galileons} and k-\emph{essence}.
The fluid variables not only have a clear physical meaning but also
drastically simplify the analysis of the system. The fluid carries
charges corresponding to shifts in field space. This shift-charge
current contains a spatial part responsible for diffusion of the charges.
Moreover, in the incompressible limit, the equation of motion becomes
the standard diffusion equation. The fluid is indeed imperfect because
the energy flows neither along the field gradient nor along the shift
current. The fluid has zero vorticity and is not dissipative: there
is no entropy production, the energy-momentum is exactly conserved,
the temperature vanishes and there is no shear viscosity. Still, in
an expansion around a perfect fluid one can identify terms which correct
the pressure in the manner of bulk viscosity. We close by formulating
the non-trivial conditions for the thermodynamic equilibrium of this
imperfect fluid. 
}
\maketitle

\section{Introduction}

Scalar field theories with non-canonical derivative interactions,
often referred to as \mbox{k-\emph{essence}} \cite{ArmendarizPicon:2000ah,ArmendarizPicon:2000dh}, have gained significant attention in the past decade as novel and
phenomenologically different models for the inflationary stage of
the early universe \cite{ArmendarizPicon:1999rj,Garriga:1999vw} and of the recent universe, dominated by dark energy
(``DE''), see \cite{ArmendarizPicon:2000ah,ArmendarizPicon:2000dh,ArkaniHamed:2003uy} and dark matter (``DM''), see e.g. Refs \cite{Scherrer:2004au,Giannakis:2005kr,ArmendarizPicon:2005nz,Bertacca:2007ux,Creminelli:2008wc,Creminelli:2009mu,Bertacca:2010ct,Furukawa:2010gr}.
The key new observable appearing in these theories in comparison to
canonical scalar fields is the speed of sound which can be very different
from the speed of light. 

On the other hand, modifications of gravity like \cite{Dvali:2000hr,Gabadadze:2009ja} in the so-called decoupling limit \cite{lpr,Nicolis:2004qq,Gabadadze:2006tf} \footnote{For more recent development, see  \cite{Babichev:2009ee,Babichev:2009jt}} reduce to non-canonical scalar field theory structurally different from \mbox{k-\emph{essence}}.  In particular, in Minkowski spacetime this scalar field theory by construction possesses  the \emph{galilean} symmetry: $\partial_{\mu}\phi\rightarrow \partial_{\mu}\phi+c_{\mu}$  where $c_{\mu}=\text{const}$. A \emph{galilean}-symmetric generalisation of this scalar field theory was introduced in  \cite{Nicolis:2008in,Nicolis:2009qm}\footnote{For earlier works unrelated to the decoupling limit, see e.g.  \cite{Fairlie:1991qe,Fairlie:1992nb,Fairlie:2011md}}. Further, it was found that it is impossible to covariantise these models in such a way so as to maintain the \emph{galilean} symmetry for dynamical spacetimes in a self-consistent manner \cite{covGal,Deffayet:2009mn}.
Nonetheless, these theories have been named \textit{galileon} theories and a significant
body of work has already been carried out analysing their impact not
only on cosmological solutions \cite{Chow:2009fm,Silva:2009km,Kobayashi:2009wr,DeFelice:2010jn,DeFelice:2010gb,Creminelli:2010ba,DeFelice:2010pv,Gannouji:2010au,Kobayashi:2010wa,DeFelice:2010nf,Mizuno:2010ag,Burrage:2010cu,Mota:2010bs,Nesseris:2010pc,Creminelli:2010qf,DeFelice:2011zh,DeFelice:2010as,Wyman:2011mp},
but also in other circumstances \cite{Hinterbichler:2010xn,Padilla:2010de,Padilla:2010ir,Padilla:2010tj,Babichev:2010kj,VanAcoleyen:2011mj,deRham:2010eu,Deffayet:2011gz,Andrews:2010km,Goon:2010xh}.
In \cite{Deffayet:2010qz}, we introduced a class of models which
extends both k-\emph{essence} and the action of the DGP \cite{Dvali:2000hr} decoupling
limit by\[
S_{\phi}=\int\mathrm{d}^{4}x\:\sqrt{-g}\left[K(\phi,\partial\phi)+G(\phi,\partial\phi)\Box\phi\right]\,,\label{eq:act11}\]
which we have utilised as an \emph{Imperfect Dark Energy} (see also
Ref. \cite{Kobayashi:2010cm}, where this class of models was slightly later
studied under the name \emph{G-inflation}).\footnote{For a further generalisation of galileons and our action \eqref{eq:act11} see Ref.~\cite{Deffayet:2011gz}} This action does not possess
the \emph{galilean} symmetry and yet it does not contain a new degree of
freedom, even in the presence of gravity. We have
found a number of surprising features exhibited by this class of models.
For example, contrary to k-\textit{essence} and the \textit{ghost condensate}
\cite{ArkaniHamed:2003uy}, even in the limit where the scalar is
an exact Goldstone boson, without direct couplings to matter, there
exist attractors in expanding cosmologies in which the scalar field
monitors and responds to the external energy density, only to eventually
arrive at a final de-Sitter state; the null energy condition is generically
violated in a stable manner and the system can evolve so as to cross
the phantom divide. These features arise as a consequence of what
we have named \emph{kinetic gravity braiding}, the essential mixing
of the derivatives of the scalar and of the metric, which cannot be
undone by a field redefinition and which necessarily modifies gravity.
The result of this \emph{kinetic braiding} is that the energy-momentum
tensor can no longer be brought to the perfect-fluid form. A number
of works have followed up the analysis, investigating structure formation
and inflation in this class of models \cite{Kimura:2010di,Kamada:2010qe,Gong:2011uw,Kobayashi:2011pc,Naruko:2011zk}.

In this work we extend our heretofore investigation of the cosmological
solutions to generic backgrounds and analyse the models with \emph{kinetic
gravity braiding} as fluids. This turns out to be a very fruitful
framework in which to proceed and we demonstrate that it results in
a very significant simplification of the equations and in giving a
very physical picture to the a priori complex dynamics of the system.
For the complete list of fluid variables and notation see table \ref{tab:comp}.

We begin by recapping parts of our discussion on the equations of
motion already presented in Ref. \cite{Deffayet:2010qz} in section \ref{sec:Action}.
Then, in section \ref{sec:Imperfect}, we show that a fluid description
is still possible. Just as in the case of k-\emph{essence}, one identifies
the time-like gradient of the scalar field with the fluid's velocity
in this way choosing a local rest frame. This implies that the scalar
field plays the role of the \textit{internal clock} for the fluid.
Since the shift-symmetric Lagrangian is necessarily asymmetric with
respect to $\phi\rightarrow-\phi$ there is a built-in \textit{arrow
of time}. Secondly one can realise that the equation of motion for
the scalar takes the form of a divergence of a current, which is conserved
in the case of a shift-symmetric Lagrangian. This allows us to identify
the shift-charge density, $n$, in the local rest frame. What will
prove key in the discussion here, is that this frame is neither the
frame in which the charges are at rest, nor is it the frame in which
there is no energy flux. However, it is in this particular frame that
vorticity vanishes providing the hypersurface of constant intrinsic
time on which the Cauchy data can be posed.

We follow Schutz \cite{Schutz} and interpret the derivative of the
scalar field with respect to proper time as the chemical potential,
$m$. The surprising new feature is that both the charge density and
the the energy density now explicitly contain the expansion of the
fluid elements. In addition to its thermodynamical part $P\left(m\right)$,
the pressure, $\mathcal{P}$, contains a term proportional to the
first time derivative of the chemical potential. As we will show,
these terms are a signature of the fluid's generically being out of
thermodynamical equilibrium. In addition, this fluid is now
\emph{imperfect}. Indeed, the energy flows neither along the field
gradient nor along the shift current.%
\footnote{Note the this definition of imperfection is different from the one
used in Ref. \cite{Unnikrishnan:2010ag,Arroja:2010wy}.%
} Therefore the energy-momentum tensor contains off-diagonal energy-flow terms. This energy flow is, in fact, diffusion occurring along
gradients of the chemical potential, \[
\mathbf{q}=-\kappa\mathbf{\boldsymbol{\nabla}}m\,,\]
where $\mathbf{\boldsymbol{\nabla}}$ is the spatial gradient
and $\kappa$ is the \emph{diffusivity}. This is a dissipationless
form of diffusion, with no entropy production. The fluid in this picture
has zero temperature, is vorticity-free and does not have any shear
viscosity, just one would expect of a system described by a single
scalar field and an action. Thus it is natural to identify this fluid with an imperfect superfluid.
The use of the fluid variables allows
us to understand the dynamics described by the highly nonlinear Amp\`{e}re-Monge-like
equation of motion as the simple first law of thermodynamics \[
\mbox{d}E=-\mathcal{P}\mbox{d}V+m\mbox{d}\mathcal{N}\,,\]
where $V$ is the comoving volume and $\mathcal{N}$ is the number
of shift charges contained in $V$. In particular, the function $K$ in the Lagrangian \eqref{eq:act11} provides the equation of state for the fluid, while the function $G$ describes the dependence of the diffusivity on the chemical potential. Note that the usual Galileon choice for the function $G=(\partial\phi)^2$ is not necessarily one which is physically motivated in this fluid picture, as we discuss in section~\ref{sec:How-unnaturally-complicated}.

In section \ref{sec:Diffusion}, we show that when the fluid dynamics
are expanded around the perfect fluid, with the diffusivity $\kappa$
employed as the expansion parameter, one obtains a correction to the
pressure that is proportional to the expansion of the fluid elements,
and therefore behaves in the manner of bulk viscosity. However, when
a proper gradient expansion is taken, we show that this bulk viscosity
disappears, confirming that there is no dissipation. That being said,
the fluid does respond to the expansion, and therefore we propose
that it is viscid in this generalised, non-dissipative sense. Secondly,
in the incompressible limit, the equation of motion for the scalar
field can be re-expressed as the diffusion equation for charge density\[
\dot{n}=-\mathbf{\boldsymbol{\nabla}}\left(\mathfrak{D}\mathbf{\boldsymbol{\nabla}}n\right)+\ldots\]
with the diffusion coefficient $\mathfrak{D}\equiv-\kappa\left(m\partial n/\partial m\right)^{-1}$,
where we have ignored the terms arising from the non-inertial nature
of the rest frame. Yet again we reiterate: this diffusion process
is dissipationless and occurs at zero temperature.

Using the fluid language for k-\emph{essence,} we show that what is
usually considered to be the relation between pressure and energy
density (the ``equation of state'') is in fact nothing but the Euler
relation for the thermodynamical representation of the fluid, relating
the energy and pressure with the particle number and chemical potential.
This relation guarantees the conservation of momentum. Surprisingly,
the Euler relation is still valid for our fluid, but one must replace
the total pressure with only its thermodynamical part. 

\medskip{}

\begin{table}[H]
\caption{Comparison of properties and notation of k-\emph{essence} with the
imperfect fluid of \emph{kinetic gravity braiding} in the shift-symmetric
case.\label{tab:comp}}
\begin{tabular}{llrrr}
\hline \hline
\textsf{\textbf{Description}} & \textsf{\textbf{Notation}} &  &  & \textsf{\textbf{Definition$\qquad\qquad$}}\tabularnewline
\hline
\hline 
\noalign{\vskip\doublerulesep}
Standard Kinetic Term & $X$ & \multicolumn{3}{r}{$ $$\frac{1}{2}g^{\mu\nu}\nabla_{\mu}\phi\nabla_{\nu}\phi\qquad\qquad$}\tabularnewline[\doublerulesep]
\hline 
\noalign{\vskip\doublerulesep}
Fluid Velocity & $u_{\mu}$ & \multicolumn{3}{r}{$\nabla_{\mu}\phi/\sqrt{2X}\qquad\qquad$}\tabularnewline[\doublerulesep]
\hline 
\noalign{\vskip\doublerulesep}
Chemical Potential & $m$ & \multicolumn{3}{r}{$\sqrt{2X}\qquad\qquad$}\tabularnewline[\doublerulesep]
\hline 
\noalign{\vskip\doublerulesep}
Expansion & $\theta$ & \multicolumn{3}{r}{$\nabla_{\mu}u^{\mu}\qquad\qquad$}\tabularnewline[\doublerulesep]
\hline 
\noalign{\vskip\doublerulesep}
Spatial Projector & $\perp_{\mu\nu}$ & \multicolumn{3}{r}{$g_{\mu\nu}-u_{\mu}u_{\nu}\qquad\qquad$}\tabularnewline[\doublerulesep]
\hline 
\noalign{\vskip\doublerulesep}
Time Derivative of $x$ & $\dot{x}$$ $ & \multicolumn{3}{r}{$u^{\mu}\nabla_{\mu}x\qquad\qquad$}\tabularnewline[\doublerulesep]
\hline 
\hline 
\noalign{\vskip\doublerulesep}
\textsf{\textbf{Description}} & \textsf{\textbf{Notation}} & \textsf{\textbf{k-}}\textsf{\textbf{\emph{Essence}}} &  & \textsf{\textbf{\emph{Kinetic Gravity Braiding}}}\textsf{\textbf{ }}\tabularnewline[\doublerulesep]
\hline \hline
\noalign{\vskip\doublerulesep}
Lagrangian & $\mathcal{L}$ & $K\left(X\right)$ &  & $K\left(X\right)+G\left(X\right)\Box$$\phi$\tabularnewline[\doublerulesep]
\hline 
\noalign{\vskip\doublerulesep}
Diffusivity & $\kappa$ & $0$ &  & $2XG_{X}$\tabularnewline[\doublerulesep]
\hline 
\noalign{\vskip\doublerulesep}
Thermal Pressure & $P$ & $K$ &  & $K$\tabularnewline[\doublerulesep]
\hline 
\noalign{\vskip\doublerulesep}
Charge Density & $n$ & $\partial P/\partial m$ &  & $\partial P/\partial m+\kappa\theta$\tabularnewline[\doublerulesep]
\hline 
\noalign{\vskip\doublerulesep}
Charge Current & $J_{\mu}$ & $nu_{\mu}$ &  & $nu_{\mu}-\left(\kappa/m\right)\perp_{\mu}^{\nu}\nabla_{\nu}m$\tabularnewline[\doublerulesep]
\hline 
\noalign{\vskip\doublerulesep}
Total Pressure & $\mathcal{P}$ & $P$ &  & $P-\kappa\dot{m}$\tabularnewline[\doublerulesep]
\hline 
\noalign{\vskip\doublerulesep}
Total Energy Density & $\mathcal{E}$ & $mn-P$ &  & $mn-P$\tabularnewline[\doublerulesep]
\hline 
\noalign{\vskip\doublerulesep}
Energy Flow & $q_{\mu}$ & $0$ &  & $q_{\mu}=-\kappa\perp_{\mu}^{\nu}\nabla_{\nu}m$\tabularnewline[\doublerulesep]
\hline 
\noalign{\vskip\doublerulesep}
Diffusion Coefficient & $\mathfrak{D}$ & $0$ &  & $-\kappa\left(m\partial n/\partial m\right)^{-1}$\tabularnewline[\doublerulesep]
\hline 
\noalign{\vskip\doublerulesep}
Force Density & $f_{\mu}$ & $\perp_{\mu}^{\nu}\nabla_{\nu}P$ &  & $\perp_{\mu}^{\lambda}\nabla_{\lambda}P+\kappa\theta\perp_{\mu}^{\lambda}\nabla_{\lambda}m$\tabularnewline[\doublerulesep]
\hline 
\noalign{\vskip\doublerulesep}
Energy Conservation & $u_{\nu}\nabla_{\mu}T^{\mu\nu}=0$ & $ $$\mbox{d}E=-\mathcal{P}\mbox{d}V$ &  & $\mbox{d}E=-\mathcal{P}\mbox{d}V+m\mbox{d}\mathcal{N}$\tabularnewline[\doublerulesep]
\hline\hline
\end{tabular}

\end{table}

In section \ref{sec:Equilibrium}, we turn to a discussion of the
conditions required for thermodynamic equilibrium in our \textit{braided superfluid}. A new result here (which also applies to k-\emph{essence}, i.e.\ the usual perfect superfluid) is that, assuming our fluid interpretation, thermodynamic equilibrium
implies the so-called Born-rigidity condition: the expansion and shear
vanish; it turns out that the configuration must also be static. However,
the key difference between these configurations and those considered
previously (see e.g., \cite{ArmendarizPicon:2005nz}) is that the
gradient of the scalar field remains time-like: therefore these type
of equilibrium configurations only exist in theories with shift-symmetric
Lagrangians. This also directly follows from our interpretation of
$\phi$ as the\textit{ intrinsic clock}.

A new feature arising from \emph{kinetic gravity braiding}
is that the equation of motion for the scalar field provides a new
condition which must be fulfilled by equilibrium configurations. In
particular, this condition constrains the equation of state which
the scalar must adopt. In the extreme case of a constant diffusivity
$\kappa$, the total equation of state for the braided scalar together
with external matter must be $w=-1/3$. In equilibrium, the scalar
must arrange itself in such a way so as to screen the gravitational
effects of itself and any external mass present in the system. 

We close by describing the motion of degenerate relativistic fermions
by a k-\emph{essence }scalar-field theory in \ref{sec:How-unnaturally-complicated}.
While the former is a simple physical system, the k-\emph{essence
} describing it appears highly technically unnatural. 

In table \ref{tab:comp}, we are providing a quick dictionary defining
our terms and contrasting the hydrodynamical properties of k-\emph{essence}
and the fluid with \emph{kinetic gravity braiding. }We assume the
shift-symmetric case where the analogy with hydrodynamics is exact.

\section{Action and Equations of Motion}%
\label{sec:Action}

In this section, we recap the discussion presented in our work \cite[\S2]{Deffayet:2010qz} where we introduced a class of scalar field theories minimally
coupled to gravity, described by the action%
\footnote{we use the metric signature convention $\left(+-\,-\,-\right)$.%
} \begin{equation}
S_{\phi}=\int\mbox{d}^{4}x\sqrt{-g}\left[K+G\times B\right],\label{e:action}\end{equation}
where we have denoted \begin{eqnarray*}
X\equiv\frac{1}{2}g^{\mu\nu}\nabla_{\mu}\phi\nabla_{\nu}\phi, & \text{and} & B\equiv g^{\mu\nu}\nabla_{\mu}\nabla_{\nu}\phi\equiv\Box\phi\,,\end{eqnarray*}
and $K\left(\phi,X\right)$ and $G\left(\phi,X\right)$ are arbitrary
functions of the scalar field and its standard kinetic term $X$.
Here and in most of the paper, we use reduced Planck units where $M_{\text{Pl}}=\left(8\pi G_{\text{N}}\right)^{-1/2}=1$.
For further discussion, it is convenient to use the standard notation
for the corresponding Lagrangian \[
\mathcal{L}\left(\phi,X,B\right)=K\left(\phi,X\right)+G\left(\phi,X\right)\times B\,,\]
which we consider as a function of three \emph{independent }variables
$\phi$, $X$ and $B$. One should note that, if the system is symmetric
with respect to constant shifts in field space: $\phi\rightarrow\phi+c$,
then the Lagrangian is \emph{necessarily not symmetric} with respect
to $\phi\rightarrow-\phi$. 

The Lagrangian $\mathcal{P}$ equivalent to $\mathcal{L}$ can be
obtained by integrating by parts the scalar-field contribution $S_{\phi}$
of the action Eq.$\,$\eqref{e:action}: \begin{equation}
\mathcal{P}=K-\left[\left(\nabla^{\lambda}\phi\right)\nabla_{\lambda}\right]G=K-2XG_{\phi}-G_{X}\nabla^{\lambda}\phi\nabla_{\lambda}X\,,\label{e:L2}\end{equation}
where the subscripts $\phi$ and $X$ denote partial differentiation
with respect to these independent variables. It is clear that it is
the dependence of $G$ on the field's gradient, $X$, that prevents
this theory from being recast in a k-\emph{essence} form, whereas
the derivative of $G$ with respect to $\phi$ contributes as an additional
k-\emph{essence} term. 

In Ref. \cite{Deffayet:2010qz}, we showed that the equation of motion
for the scalar field takes the following form 
\begin{equation}
\nabla_{\mu}J^{\mu}=\mathcal{P_{\phi}}\,,
\label{eq:DivJeom}
\end{equation}
where $\mathcal{P}_{\phi}=K_{\phi}-\left[\left(\nabla^{\lambda}\phi\right)\nabla_{\lambda}\right]G_{\phi}$
and the current \begin{equation}
J_{\mu}=\left(\mathcal{L}_{X}-2G_{\phi}\right)\nabla_{\mu}\phi-G_{X}\nabla_{\mu}X\,,\label{e:Jmu}\end{equation}
corresponds to the Noether current for Lagrangians invariant under
constant shifts in field space: $\phi\rightarrow\phi+c$. Thus this
current is conserved provided that $\mathcal{P_{\phi}}=0$. The fully
expanded equation of motion is somewhat unwieldy and was presented
in Ref. \cite{Deffayet:2010qz}, see Eq.~(2.5) on page 6; it will not be necessary in the discussion
here. 
We would like to stress that this second-order partial differential equation in the single scalar variable $\phi$ is the only dynamical equation besides the Einstein equations. In particular, this is the equation which one has to solve in numerical analysis.
The presence of third-order derivatives might have been expected owing
to the d'Alem\-bertian term in the action; however, as we have proven
in \cite{Deffayet:2010qz}, the equation of motion contains at most
second-order derivatives. Therefore theories with \emph{kinetic gravity
braiding} do not contain any additional hidden (and ghosty) degrees
of freedom which would appear by virtue of the Ostrogradsky procedure.
Further, for any coordinate $x^{i}$, the second derivative $\nabla_{i}\nabla_{i}\phi$
appears only linearly in the equation of motion. This allows one to
associate this equation of motion with a particular generalisation
of the Amp\`{e}re-Monge equation for four-dimensional manifolds with
Lorentzian signature. In section \ref{sub:EoM}, we present a different
derivation of the expanded equation of motion which makes use of the
Raychaudhuri equation and explicitly shows how the higher-order derivatives
are eliminated.

The energy-momentum tensor (``EMT'') for the scalar field is most
easily derived in the standard way using the Lagrangian presented
in Eq.$\,$\eqref{e:L2}\begin{equation}
T_{\mu\nu}\equiv\frac{2}{\sqrt{-g}}\frac{\delta S_{\phi}}{\delta g^{\mu\nu}}=\mathcal{L}_{X}\nabla_{\mu}\phi\nabla_{\nu}\phi-g_{\mu\nu}\mathcal{P}-\nabla_{\mu}G\nabla_{\nu}\phi-\nabla_{\nu}G\nabla_{\mu}\phi\,.\label{e:EMT}\end{equation}

It is key to observe that the EMT also contains second derivatives
of $\phi$ appearing in $\mathcal{P}$, $\mathcal{L}_{X}$ and $\nabla_{\mu}G$.
Therefore as we have already discussed in Ref. \cite{Deffayet:2010qz},
both the Einstein's equations as well as the equation of motion for
the scalar field \eqref{eq:DivJeom} contain second derivatives of
both the metric $g_{\mu\nu}$ and the scalar field $\phi$, so that
the system is not diagonal in the second derivatives and no field
redefinition exists which would unmix the two fields. It is this essential
kinetic mixing of gravity and the scalar that we have named \emph{kinetic
gravity braiding. }

\section{Fluid Picture}%
\label{sec:Imperfect}

When considering scalar-field theories as applied to cosmology, one
usually restricts to the case with timelike field derivatives. It
is well known that many aspects of k-\emph{essence} in such a setup
can be described in terms of relativistic hydrodynamics \cite{ArmendarizPicon:1999rj}.
Below we introduce a fluid picture for the theory given by the action
\eqref{e:action}. As in the case of k-\emph{essence}, this hydrodynamical/fluid
language turns out to be helpful in developing intuition and in simplifying
notation. To our best knowledge the analysis presented below appears
in the literature for the first time. For earlier studies of thermodynamics of relativistic potential flows and k-\emph{essence} see e.g.\ Refs \cite{Landafshitz_Fluid,Moncrief:1980,DiezTejedor:2005fz,Bilic:2008zk}.

\subsection{Four-Velocity and Kinematical Decomposition}%
\label{sec:KinDecomposition}

First of all, by analogy with k-\emph{essence} we can introduce an
effective four-velocity \begin{equation}
u_{\mu}\equiv\frac{\nabla_{\mu}\phi}{\sqrt{2X}}\,,\label{eq:4velocity}\end{equation}
which defines the local rest frame (LRF). We will then use the notation
\[
\dot{\left(\;\;\right)}=\frac{\mbox{d}}{\mbox{d}\tau}\left(\;\;\right)=u^{\alpha}\nabla_{\alpha}\left(\;\;\right)\,,\]
for the derivative along $u^{\alpha}$ (material derivative), thus
making $\tau$ the proper time of an observer comoving with the LRF.
We can then introduce the transverse projector \begin{equation}
\perp_{\mu\nu}=g_{\mu\nu}-u_{\mu}u_{\nu}\,,\label{eq:Projector}\end{equation}
which will be the first fundamental form in the hypersurfaces $\Sigma_{\phi}:\phi(x)=\text{const}$
and which will allow us to decompose vectors (and gradients) into
time-like and space-like parts as observed in the LRF,\[
\nabla_{\mu}=u_{\mu}u^{\lambda}\nabla_{\lambda}+\perp_{\mu}^{\lambda}\nabla_{\lambda}=u_{\mu}\frac{\mbox{d}}{\mbox{d}\tau}+\mathbf{\boldsymbol{\overline{\nabla}}}_{\mu}\,.\]
Without loss of generality, we will assume that the time derivative
of $\phi$ is positive definite, always making $u^{\mu}$ future directed.
The field $\phi$ can thus be considered to be an \emph{internal clock}.
The hypersurfaces $\Sigma_{\phi}$ then are Cauchy hypersurfaces for
the standard-model fields and gravity and natural candidates for being
the Cauchy hypersurfaces for the equation of motion of the field $\phi$
(see the discussion regarding this particular choice of frame in appendix
\ref{sub:frames}). The equation of motion \eqref{eq:DivJeom} is
of the second order, therefore from the Cauchy-Kowalewski theorem,
the initial data on $\Sigma_{\phi}$ are $\dot{\phi}_{0}\left(\boldsymbol{x}\right)$
and $\phi_{0}\left(\boldsymbol{x}\right)$, which is represented not
only by its value ($\phi_{0}\left(\boldsymbol{x}\right)=\text{const}$
on $\Sigma_{\phi}$) but also by second derivatives with respect to
the spatial coordinates $\boldsymbol{x}$ on $\Sigma_{\phi}$. 

In the shift-symmetric case, the Lagrangian is explicitly independent
of the \emph{internal clock $\phi$. }Then, as we have already mentioned,
the Lagrangian necessarily violates $\phi\rightarrow-\phi$ symmetry
furnishing our model with a built-in \emph{arrow of time} or an asymmetry
of the motions with respect to an increasing or decreasing \emph{internal
clock} $\phi$. 

The corresponding four-acceleration is defined as \begin{equation}
a_{\mu}\equiv\dot{u}_{\mu}\equiv u^{\lambda}\nabla_{\lambda}u_{\mu}\,.\label{eq:4acceleration}\end{equation}
 and can also be rewritten in form of a spatial gradient, \begin{equation}
a_{\mu}=\perp_{\mu}^{\lambda}\nabla_{\lambda}\ln\dot{\phi}\,,\label{eq:AasGradient}\end{equation}
It will be convenient to make use of the notation for its magnitude,
\begin{equation}
a\equiv\sqrt{-a^{\lambda}a_{\lambda}}\,.\label{aDeff}\end{equation}
Further, we can make the standard kinematical decomposition,\begin{equation}
\nabla_{\mu}u_{\nu}=u_{\mu}a_{\nu}+\frac{1}{3}\theta\perp_{\mu\nu}+\sigma_{\mu\nu}\,,\label{eq:Decomposition}\end{equation}
where $\theta$ denotes the expansion, \begin{equation}
\theta\equiv\perp_{\mu}^{\lambda}\nabla_{\lambda}u^{\mu}=\nabla_{\mu}u^{\mu}\,,\label{eq:Expansion}\end{equation}
and $\sigma_{\mu\nu}$ is the shear tensor, \begin{equation}
\sigma_{\mu\nu}\equiv\frac{1}{2}\left(\perp_{\mu}^{\lambda}\nabla_{\lambda}u_{\nu}+\perp_{\nu}^{\lambda}\nabla_{\lambda}u_{\mu}\right)-\frac{1}{3}\perp_{\mu\nu}\theta\,,\label{eq:shear}\end{equation}
which is symmetric, traceless and purely spatial. In this decomposition,
we have used the vanishing of the rotation tensor (or twist) $\perp_{\mu}^{\lambda}\nabla_{\lambda}u_{\nu}-\perp_{\nu}^{\lambda}\nabla_{\lambda}u_{\mu}$
in accordance with the Frobenius theorem. The four-acceleration $a_{\mu}$,
expansion $\theta$ and the shear tensor $\sigma_{\mu\nu}$ are constructed
from spatial derivatives of the scalar field and can be calculated
on any $\Sigma_{\phi}$ directly from the initial data without using
the equation of motion. Finally, it is useful to introduce the extrinsic
curvature of the hypersurface $\Sigma_{\phi}$, 

\begin{equation}
\mathcal{K}_{\alpha\beta}=\perp_{\alpha}^{\gamma}\perp_{\beta}^{\lambda}\nabla_{\lambda}u_{\gamma}=\sigma_{\alpha\beta}+\frac{1}{3}\perp_{\alpha\beta}\theta\,,\label{Extrisic}\end{equation}
from whence it follows that $\mathcal{K}=\theta$.

\subsection{Shift-Charge or Particle Current}

Using the formulae \begin{eqnarray}
\nabla_{\mu}X=2Xa_{\mu}+\dot{X}u_{\mu}\,, & \mbox{and} & \Box\phi=\frac{\dot{X}}{\sqrt{2X}}+\sqrt{2X}\theta\,,\label{eq:decomposition}\end{eqnarray}
we can write the Noether current Eq. \eqref{e:Jmu}, conserved only
when shift symmetry is present, as \begin{equation}
J_{\mu}=\left(\dot{\phi}\left(K_{X}-2G_{\phi}\right)+\kappa\theta\right)u_{\mu}-\kappa a_{\mu}\,,\label{eq:currentAU}\end{equation}
where we have introduced the notation%
\footnote{Note that in Ref. \cite[Eq. (3.6)]{Deffayet:2010qz}, we used
a definition for $\kappa$ that is smaller than this one by a factor
of 2, $\kappa_{\text{here}}=2\kappa_{\text{there}}$. This served
to simplify the equations even more in the particular case of a cosmological
background, but the proper normalisation for this term which gives
its physical meaning is the one presented in this work. %
} \begin{equation}
\kappa\equiv2XG_{X}\,.\label{eq:diffusionCoeff}\end{equation}
From the standard discussion of imperfect fluids (see e.g. Ref. \cite{Andersson:2006nr}),
we have to identify the space-like part of the current \eqref{eq:currentAU},
$-\kappa a_{\mu}$, with \emph{diffusion,} while $\kappa$ can be
interpreted as the \emph{diffusivity}, which we will take
to be the coefficient relating the energy flow to the gradient of
the chemical potential%
\footnote{Note that usually the terms \emph{diffusivity }and \emph{diffusion
coefficient }are used interchangeably, since the same coefficient
usually appears in both the Fick's laws. We have kept the definition
of diffusivity as the one here and use it to mean a coefficient which
determines an energy flux rather than a charge flux: in our case,
this is a more natural definition. In section \ref{sec:Diffusion}
we show that, in an expansion around an incompressible fluid, the
equation of motion reduces to the diffusion equation with the diffusion
coefficient related to, but not the same as, the diffusivity, see
Eq. \eqref{eq:DiffusionProper}.%
} (see Eq. \eqref{EnergyFlowDiffusion}). Despite the presence of second
derivatives in \eqref{e:Jmu}, the current does not contain the second
time derivative of the field $\phi$ and can be calculated directly
from initial data. The density of the shift-symmetry charge%
\footnote{Note that in Ref. \cite[Eq. (3.2)]{Deffayet:2010qz}, we denoted this
number density by $J$: $n_{\text{here}}=J_{\text{there}}$.%
} \begin{equation}
n\equiv u^{\mu}J_{\mu}=\dot{\phi}\left(K_{X}-2G_{\phi}\right)+\kappa\theta\,,\label{eq:particle density}\end{equation}
contains the expansion $\theta$, which is highly non-standard.
It is convenient to associate this Noether current with the particle-number
current. In such a way, the introduced total particle/charge number
will be conserved provided there be shift symmetry. However, the number
of particles $\mathcal{N}\equiv nV$ in an infinitesimal volume $V$
moving with velocity $u^{\mu}$ (comoving volume) is not conserved.
Indeed using the decomposition \eqref{eq:currentAU} one can write
equation of motion \eqref{eq:DivJeom} in the form \begin{equation}
\dot{n}+\theta n-\nabla_{\mu}\left(\kappa a^{\mu}\right)=\mathcal{P}_{\phi}\,,\label{eq:eomDiffusionDiv}\end{equation}
and using the geometric meaning of the expansion \begin{equation}
\theta=\frac{\dot{V}}{V}\,,\label{eq:ExpansionVsVolume}\end{equation}
 we can obtain \begin{equation}
\dot{\mathcal{N}}=V\left(\mathcal{P}_{\phi}+\nabla_{\mu}\left(\kappa a^{\mu}\right)\right)=V\mathcal{P}_{\phi}+\dot{\mathcal{N}}_{\text{dif}}\,,\label{eq:ParticleProdRate}\end{equation}
where $\dot{\mathcal{N}}_{\text{dif}}$ denotes the number of particles
transported to the volume by diffusion per unit of proper time. Thus
generically $\mathcal{N}$ is not conserved even in the shift-symmetric
case. The total particle current is subluminal provided\begin{equation}
n^{2}-\kappa^{2}a^{2}\geq0\,,\label{EckartExists}\end{equation}
If Eq. \eqref{EckartExists} holds, then one can chose an alternative
fluid description where one takes the Eckart frame which moves with
the particles as the LRF. We will comment more on this later.

\subsection{Chemical Potential and Force }

Using the general expression for the Energy-Momentum Tensor (``EMT'')
\eqref{e:EMT} and the decomposition \eqref{eq:Decomposition}, we
define the energy density of the fluid in the standard way \begin{equation}
\mathcal{E}\equiv T_{\mu\nu}u^{\mu}u^{\nu}=2X\left(K_{X}-G_{\phi}\right)+\kappa\theta\dot{\phi}-K\,.\label{EnergyDensity}\end{equation}
Contrary to the usual descriptions of relativistic imperfect fluids,
the energy density contains the expansion, $\theta$. 

For the shift-symmetric case we recall the definition of the effective
mass per charge (which is also the chemical potential or injection
energy) see e.g. \cite[p. 562]{MTW} and obtain \[
m\equiv\left(\frac{\partial\mathcal{E}}{\partial n}\right)_{\tau,V=\text{const}}=\left(\frac{\partial\mathcal{E}}{\partial n}\right)_{\phi=\text{const},\theta=0}=\dot{\phi}\,,\]
this expression is identical to that of Ref. \cite{Schutz}. Further,
we will identify \begin{equation}
m\equiv\dot{\phi}=\sqrt{2X}>0\,,\label{eq:effectiveM}\end{equation}
with an effective rest mass per particle building the fluid even in
non-shift-symmetric cases. As we will see later (see Eqs \eqref{eq:Euler},
\eqref{eq:FirstLaw}) the rest-mass introduced in this way plays the
role of the chemical potential.%
\footnote{In the literature, $m$ is also called specific enthalpy or Gibbs
energy per particle. %
} 

The effective momentum of the particle can then be defined as \begin{equation}
p_{\mu}\equiv mu_{\mu},\label{eq:momentum}\end{equation}
and the corresponding relativistic vorticity, $\nabla_{\mu}p_{\nu}-\nabla_{\nu}p_{\mu}$,
is zero.%
\footnote{Here it is worth recalling that in a stationary gravitational field
with Killing vector $\partial_{t}$ the conserved energy of a particle
is governed by $p_{t}.$ Using our identification we obtain that $p_{t}=\partial_{t}\phi$
has to be conserved. As it was shown for k-\emph{essence }in \cite{Akhoury:2008nn},
the energy $p_{t}$ given by \eqref{eq:momentum} is indeed conserved
for stationary configurations. Taking into account that $\phi$ plays
the role of an \emph{internal clock} one can expect that stationarity
implies shift symmetry in field space. Thus we conjecture that the
statement from \cite{Akhoury:2008nn} is also applicable to the \emph{kinetic
gravity braiding} theories. %
} Further, the fact that $p_{\mu}p^{\mu}=m^{2}$ and $p_{\mu}=\nabla_{\mu}\phi$,
means that $-\phi$ plays the role of Hamilton's principal function
(action as a function of final coordinates) for each shift-charge.
Indeed, for the action of a particle we have \[
s=-\int md\tau=-\phi\,,\]
where we have used Eq. \eqref{eq:effectiveM} and where we have omitted
the constant of integration. 

Exploiting the particle momentum, one can introduce a relativistic
3-force $F_{\mu}$ acting on each particle, \begin{equation}
F_{\mu}\equiv\perp_{\mu}^{\lambda}\dot{p}_{\lambda}=\perp_{\mu}^{\lambda}\frac{\mbox{d}}{\mbox{d}\tau}\left(mu_{\lambda}\right)\,.\label{Force}\end{equation}
Then, using the decomposition \begin{equation}
\nabla_{\mu}m=ma_{\mu}+\dot{m}u_{\mu}\,,\label{eq:Mgrad}\end{equation}
which follows from the definition of the effective mass \eqref{eq:effectiveM},
and decomposition \eqref{eq:Decomposition}, the force can be written
as the gradient of the chemical potential: \begin{equation}
F_{\mu}=ma_{\mu}=\perp_{\mu}^{\lambda}\nabla_{\lambda}m\,.\label{Force_Grad}\end{equation}
Here it is worth remembering that, in a system in thermodynamical
equilibrium which experiences an external potential, the chemical
potential is given by this external potential plus some constant,
so that the force acting on the particle is exactly the gradient of
the external potential, as expected (do not forget the signature).
We discuss equilibrium configurations in detail in section \ref{sec:Equilibrium}. 

Further we can also express the current Eq. \eqref{eq:currentAU}
making use of the chemical potential \begin{equation}
J_{\mu}=nu_{\mu}-\frac{\kappa}{m}\perp_{\mu}^{\lambda}\nabla_{\lambda}m\,,\label{eq:current_n}\end{equation}
c.f. with the formula (59.5) for the diffusive current from Ref. \cite[p. 231]{Landafshitz_Fluid}.

\subsection{Energy-Momentum Tensor}%
\label{sub:EMT}

In this fluid language the EMT Eq. \eqref{e:EMT} can be re-expressed
as \begin{equation}
T_{\mu\nu}=\mathcal{E}u_{\mu}u_{\nu}-\perp_{\mu\nu}\mathcal{P}+u_{\mu}q_{\nu}+u_{\nu}q_{\mu}\,,\label{eq:hydroEMT}\end{equation}
where the energy density is \begin{equation}
\mathcal{E}\equiv T_{\mu\nu}u^{\mu}u^{\nu}=2X\left(K_{X}-G_{\phi}\right)+\theta m\kappa-K\,.\label{EnergyDensity}\end{equation}
As we have already remarked, contrary to the usual descriptions of
relativistic imperfect fluids, the energy density contains the expansion,
$\theta$. Still, this expression can be calculated purely from the
initial data. 

Further, we can define the effective total isotropic pressure in the
usual fashion and find that it is given by the Lagrangian Eq. \eqref{e:L2}:
\begin{equation}
\mathcal{P}\equiv-\frac{1}{3}T^{\mu\nu}\perp_{\mu\nu}=K-2XG_{\phi}-\kappa\dot{m}\,.\label{TotalPressure}\end{equation}
The presence of the last term containing $\dot{m}$ can be associated
with an additional force transporting particles along the gradients
of the chemical potential. Because of this term, the total pressure
can be calculated only after making use of the equations of motion
which is not the case in standard fluids (see section \ref{sub:EoM}
for a further discussion). However, in configurations in thermodynamical
equilibrium this term vanishes. It will prove helpful to also introduce
the part of pressure \begin{equation}
P\equiv K-2XG_{\phi}\,,\label{EquilibriumPressure}\end{equation}
which can be directly calculated from the initial data and which does
not vanish in equilibrium configurations, usually called the thermodynamic
pressure  \footnote{This imperfection from the construction called \emph{Inhomogeneous Equation of State} where pressure is postulated to be a function of both the energy density and the Hubble parameter, see e.g. \cite{Nojiri:2005sr,Capozziello:2005pa,Nojiri:2006ri}.}.

Finally the energy flow associated with diffusion is \begin{equation}
q_{\mu}\equiv\perp_{\mu\lambda}T_{\nu}^{\lambda}u^{\nu}=-m\kappa a_{\mu}=m\perp_{\mu\nu}J^{\nu}\,.\label{eq:EnergyFlow}\end{equation}
Thus the energy flow $q_{\mu}$ arises purely from diffusive part
of the particle current $\perp_{\mu\nu}J^{\nu}$ and the heat flow,
which is defined as%
\footnote{See e.g. Ref. \cite[\S1, Eq. (39)]{deGroot}. Note that many works
use the Eckart frame as the LRF from the very beginning so that the
$q_{\mu}$ used there can only be the heat flow. %
} \[
\text{Heat Flow}\equiv\perp_{\mu\lambda}T_{\nu}^{\lambda}u^{\nu}-m\perp_{\mu\nu}J^{\nu}=0\,,\]
vanishes, as it should for a system with identically zero temperature.
Then we can see that the diffusive energy flux is proportional to
the spatial gradient of the chemical potential: \begin{equation}
q_{\mu}=-\kappa\perp_{\mu}^{\nu}\nabla_{\nu}m\,,\label{EnergyFlowDiffusion}\end{equation}
c.f. with the formula (59.5) for the diffusive energy flux from \cite[p. 231]{Landafshitz_Fluid}.
We take this relation as our definition of $\kappa$ as the \emph{diffusivity}.
It is instructive to compare these relations with formulae (285) and
(287) from the imperfect fluid from review \cite[p. 58]{Andersson:2006nr}. 

In simple equilibrium thermodynamics, the chemical potential has to
be equal throughout the body. If there are gradients of the chemical
potential, then the system develops forces and transports particles
to compensate for these gradients. As we discuss in section \ref{sec:Equilibrium},
in our case it is possible to construct equilibrium configurations
which do nonetheless exhibit diffusive flow in a time-invariant fashion.

Having defined the thermodynamic quantities, we can then re-express
the particle density Eq. \eqref{eq:particle density} \begin{equation}
n\left(\phi,m,\theta\right)=P_{m}+\kappa\theta+m\kappa_{\phi}=\mathcal{P}_{m}+\nabla_{\mu}\left(\kappa u^{\mu}\right)\,,\label{eq:nAsPm}\end{equation}
where we have explicitly stressed that we are treating the expansion
as an independent variable. This form allows us to immediately see
that $n$ can be found using solely initial data.

Using equations \eqref{eq:particle density}, \eqref{eq:effectiveM},
\eqref{EnergyDensity} and \eqref{EquilibriumPressure} we can obtain
the Euler relation:\begin{equation}
\mathcal{E}=mn-P\,.\label{eq:Euler}\end{equation}
Observe that the diffusion contribution $\kappa\dot{m}$ cancels out
and only the usual equilibrium part of the pressure, which can be
calculated from the initial data, contributes. We should stress here
that all the above results hold even when shift-symmetry is broken
and particle number is not conserved.

We would like to mention that the result Eq. \eqref{eq:Euler} in
the k-\emph{essence }limit ($G_{X}=0)$ reduces to the formula sometimes
referred to as the k-\emph{essence }equation of state, $\mathcal{E}=2XP_{X}-P$\emph{
}\cite{ArmendarizPicon:2000ah}\emph{. }To our best knowledge, the
relationship between this equation of state and the Euler relation
appears to have been missed in the literature.

\subsection{Energy-Momentum Conservation and the First Law of Thermodynamics}

It is instructive to analyse the equations resulting from the conservation
of the EMT $\nabla_{\mu}T_{\nu}^{\mu}=0$. For energy conservation
we obtain: \begin{equation}
u_{\nu}\nabla_{\mu}T^{\mu\nu}=\dot{\mathcal{E}}+\theta\left(\mathcal{E}+\mathcal{P}\right)-\nabla_{\lambda}\left(m\kappa a^{\lambda}\right)+m\kappa a_{\lambda}a^{\lambda}=0\,,\label{eq:energyConserv}\end{equation}
where we have used the definition of the expansion and the property
of four acceleration $u_{\mu}a^{\mu}=0$. Further we can express the
divergence of the diffusion current through the equation of motion
\eqref{eq:eomDiffusionDiv} \begin{equation}
\nabla_{\mu}\left(\kappa a^{\mu}\right)=m\left(\dot{n}+\theta n-\mathcal{P}_{\phi}\right)\,,\label{eq:EOM_ndot}\end{equation}
 so that the conservation of energy eventually reduces to \[
\dot{\mathcal{E}}+\theta\left(\mathcal{E}+\mathcal{P}\right)-m\left(\dot{n}+\theta n\right)+m\mathcal{P}_{\phi}=0\,.\]

Let us now consider the evolution of the energy $E\equiv\mathcal{E}V$
in the comoving volume $V$. Recalling the definition of the expansion
\eqref{eq:ExpansionVsVolume} and the ``particle conservation'' equation
\eqref{eq:ParticleProdRate} we obtain the first law of thermodynamics
for the matter in the comoving volume in the shift-symmetric case:
\begin{equation}
\mbox{d}E=-\mathcal{P}\mbox{d}V+m\mbox{d}\mathcal{N}\,,\label{eq:FirstLaw}\end{equation}
where $\mathcal{N}$ is number of particles in the volume $V$, while
$\mbox{d}$ denotes the differential along the velocity $u^{\mu}$.
From expression \eqref{eq:FirstLaw} it is obvious that $m$ plays
the role of chemical potential. In the general case with $\mathcal{P}_{\phi}\neq0$,
one should modify Eq. \eqref{eq:FirstLaw} by substituting $\mbox{d}\mathcal{N}\rightarrow\delta\mathcal{N}_{\text{dif}}$---the
infinitesimal number of particles transported to the volume by diffusion,
but not those produced by the source $\mathcal{P}_{\phi}$. 

There is only one single scalar degree of freedom $\phi$ and one
independent scalar equation of motion \eqref{eq:DivJeom}. Hence $\perp_{\mu\nu}\nabla_{\lambda}T^{\lambda\nu}=0$,
momentum conservation, cannot contain any additional information.
Indeed, one can calculate that \[
\perp_{\mu\nu}\nabla_{\lambda}T^{\lambda\nu}=\left(\mathcal{E}+P-mn\right)a_{\mu}\,,\]
therefore the Euler relation \eqref{eq:Euler} guarantees momentum
conservation.

For completeness, it is worth mentioning how the ``Archimedes law''
changes for this fluid. The force density is given by \[
f_{\mu}\equiv nF_{\mu}\,,\]
or applying the Euler relation \eqref{eq:Euler} it can be written
as \[
f_{\mu}=n\perp_{\mu}^{\lambda}\nabla_{\lambda}m=nma_{\mu}=\left(\mathcal{E}+P\right)a_{\mu}\,.\]
On the other hand, using momentum conservation $\perp_{\mu\nu}\nabla_{\lambda}T^{\lambda\nu}=0$,
after some algebra we obtain \[
f_{\mu}=\perp_{\mu}^{\lambda}\nabla_{\lambda}P+\left(\kappa_{\phi}m+\kappa\theta\right)\perp_{\mu}^{\lambda}\nabla_{\lambda}m\,.\]
Hence the force density originates from a superposition of spatial
gradients of the equilibrium part of pressure and of the chemical potential.
Using Eq. \eqref{eq:nAsPm}, we can rewrite the contents of the parentheses as $\kappa_{\phi}m+\kappa\theta=n-P_{m}$,
which disappears in thermodynamical equilibrium (it is one of the
Maxwell relations). On the other hand, in k-\emph{essence}, the last
term is not present by the simple virtue of $\kappa=0$.

\subsection{Equation of Motion}%
\label{sub:EoM}

Let us discuss the equation of motion \eqref{eq:eomDiffusionDiv}
in more detail. In particular, we will be interested in solving it
with respect to highest time derivative: $\dot{m}=\ddot{\phi}$. Naively
it seems that Eq. \eqref{eq:eomDiffusionDiv} should have third derivatives.
However, as we know from Ref. \cite{Deffayet:2010qz} these terms
with third derivatives cancel. Indeed, using Eq. \eqref{eq:Mgrad},
\eqref{TotalPressure} and the Euler integral \eqref{eq:Euler} we
obtain for Eq. \eqref{eq:eomDiffusionDiv}
\begin{equation}
\left(n_{m}+\kappa_{\phi}\right)\dot{m}+\kappa\left(\dot{\theta}-\nabla_{\mu}a^{\mu}\right)+\theta n-m\kappa_{m}a^{\lambda}a_{\lambda}+\mathcal{E}_{\phi}=0\,,\label{eq:EoM}
\end{equation}
while from the Raychaudhuri equation we have 
\begin{equation}
\dot{\theta}=-\frac{1}{3}\theta^{2}+\nabla_{\mu}a^{\mu}-\sigma_{\mu\nu}\sigma^{\mu\nu}-R_{\mu\nu}u^{\mu}u^{\nu}\,,\label{eq:Raychaudhuri}
\end{equation}
so that third derivatives disappear giving a second-order equation
\begin{equation}
\left(n_{m}+\kappa_{\phi}\right)\dot{m}-\kappa\left(\sigma_{\alpha\beta}\sigma^{\alpha\beta}+\frac{1}{3}\theta^2+R_{\mu\nu}u^{\mu}u^{\nu}\right)+\theta n-\kappa_{m}ma^{\lambda}a_{\lambda}+\mathcal{E}_{\phi}=0\,.\label{eom_with_R}\end{equation}
We then use the Gauss equation
\begin{equation}
\sigma_{\alpha\beta}\sigma^{\alpha\beta}-\frac{2}{3}\theta^2=- \phantom{}^{(3)}R-2G_{\mu\nu} u^\mu u^\nu
\end{equation}
to re-express the shear in terms of the intrinsic curvature of the spatial hypersurface and obtain for the equation of motion \eqref{eom_with_R}
\begin{equation}
\left(n_{m}+\kappa_{\phi}\right)\dot{m}+\theta n+\kappa\left(\phantom{}^{(3)}R +2G_{\mu\nu} u^\mu u^\nu -R_{\mu\nu}u^{\mu}u^{\nu}-\theta^2\right)-\kappa_{m}ma^{\lambda}a_{\lambda}+\mathcal{E}_{\phi}=0\,.
\label{eom_with_3R}
\end{equation}
This equation of motion is linear in the highest time derivative $\ddot{\phi}=\dot{m}$.
Using Einstein equations and the expression for the EMT \eqref{eq:hydroEMT},
we can eliminate the curvature term \[
2G_{\mu\nu} u^\mu u^\nu -R_{\mu\nu}u^{\mu}u^{\nu}=\frac{3}{2}\left(\mathcal{E}-P+\kappa\dot{m}\right)+W_{\text{ext}}\,,\]
where we have denoted any possible contribution to the above contraction
of the curvature tensors external to the scalar field as \begin{equation}
W_{\text{ext}}\equiv\left(T_{\text{ext}}^{\mu\nu}+\frac{1}{2}T_{\text{ext}}g^{\mu\nu}\right)u_{\mu}u_{\nu}\,.\label{eq:Wext}\end{equation}
Thus after such an elimination of the second derivatives of the metric,
the equation of motion takes the form 
\begin{equation}
\left(n_{m}+\kappa_{\phi}+\frac{3}{2}\kappa^{2}\right)\dot{m}+\theta n + \kappa\left(\phantom{}^{(3)}R+\frac{3}{2} (\mathcal{E}-P)-\theta^2\right)-\kappa_{m}ma^{\mu}a_{\mu}+\mathcal{E}_{\phi}=-\kappa W_\text{ext}\,.
\label{FluidEoM}
\end{equation}
The diffusivity $\kappa$ also serves as an effective coupling
constant in front of the external source term $W_{\text{ext}}$. Since
the total pressure $\mathcal{P}$ contains $\dot{m}$ (see Eq. \eqref{TotalPressure}),
the above equation implies that the pressure of the fluid depends
on the acceleration and the extrinsic curvature tensor (and therefore
the expansion $\theta$). Moreover, we should expect that the coefficient
in front of $\ddot{\phi}$ is responsible for the presence or absence
of ghosts / negative kinetic energies. Namely, there are no ghosts
provided \begin{equation}
D\equiv n_{m}+\kappa_{\phi}+\frac{3}{2}\kappa^{2}>0\,.\label{noGhosts}\end{equation}
We confirm this claim explicitly in our discussion of the effective
acoustic metric for perturbations in the follow-up work Ref. \cite{Metrics}.
As we have already mentioned, this condition is not correlated with
the Null Energy Condition. Using Eq. \eqref{eq:Euler} and Eq. \eqref{eq:nAsPm}
this function $D$ can also be written in the form \begin{equation}
D=\frac{\mathcal{E}_{m}-\kappa\theta}{m}+\frac{3}{2}\kappa^{2}.\label{DonE}\end{equation}
which simply reduces to the well-known k-\emph{essence }limit, $D=\mathcal{E}_{X}$
\cite{Vikman:2004dc}. Provided $D\neq0$, the solution of the Cauchy
problem exists locally under the usual limitations. On the other hand,
from the equation of motion \eqref{FluidEoM} it follows that the
zeroes of $D$ generically correspond to pole-like singularities in
$\dot{m}$. Given the dependence of the total pressure on $\dot{m}$,
Eq. \eqref{TotalPressure}, the configurations with $D=0$ are \emph{pressure
singularities} and they build a barrier in phase space which is generically
impenetrable. Thus a fluid evolving from some initial data with healthy
perturbations ($D>0$) will never become ghosty in a smooth and controllable
way. Note that, similarly to the cases studied in Ref. \cite{Vikman:2004dc},
for some Lagrangians and some external matter $W_{\text{ext}}$ there
may well exist trajectories penetrating this\emph{ }barrier\emph{.
}However, these trajectories would have measure zero in the phase
space of the system since they correspond to an exact cancellation
of the above singularity. 

Finally, we would like to stress that the fluid variables are helpful to develop intuition but do not appear to be very practical in actual numerical calculations. The equation of motion cannot be reduced to system of ordinary differential equations along a world line. In particular, the system \eqref{FluidEoM} and \eqref{eq:Raychaudhuri} is not closed and requires the somewhat unwieldy Raychaudhuri equation for the shear tensor $\sigma_{\mu\nu}$ which then has to be supplemented with an even more complicated equation for $\dot a_{\mu}$. On top of that all these equations will be subject to the constraint that the vorticity of $u^{\mu}$ vanish.  
Working with $\phi$ directly and solving the expanded \eqref{eq:DivJeom} has the advantage that there are no further constraints. Solving equations of motion for the velocity potential (phase for superfluids) $\phi$ is of course the standard practice in vorticity-free hydrodynamics.
\subsection{Diffusion and Non-Dissipative Bulk Viscosity}%
\label{sec:Diffusion}

Let us now elaborate on the diffusive interpretation of this fluid.
First of all, one quickly notices that Eq. \eqref{eq:current_n} is
a {}``1st Fick's law'' $J\propto\boldsymbol{\overline{\nabla}}m$,
relating the diffusive flux of particles in or out of a comoving volume
to the spatial gradient of the chemical potential, which again is
consistent with our interpretation of $\kappa$ as the diffusivity.
Of course, the {}``2nd Fick's law'', namely a diffusion-type equation
for the charge density $n$, can be identified with the equation of
motion. Turning the attention to the shift-symmetric case for simplicity,
we can rewrite the equation of motion \eqref{eq:EOM_ndot} as\begin{equation}
\dot{n}+\theta n=n_{m}\dot{m}+\theta n+\kappa\dot{\theta}=\nabla_{\mu}\left(\frac{\kappa}{m}\perp^{\mu\nu}\nabla_{\nu}m\right)\,,\label{eq:diff}\end{equation}
which is already a time-dependent diffusion-type equation for $m$,
with some extra terms. As we have already mentioned, the $\kappa\dot{\theta}$
term plays a crucial role to keep this equation of second order, through
the Raychaudhury equation. One could worry that at the same time this
may remove the diffusive character of Eq. \eqref{eq:diff}, but this
is actually not the case. For instance, there are configurations in
which one can have a nonzero $\nabla_{\mu}a^{\mu}$ but still $\dot{\theta}=0$
(in Section \ref{sec:Equilibrium} we discuss examples of such configurations). 

It is perhaps more illuminating to write Eq. \eqref{eq:diff} directly
in terms of $n$. For this purpose, it suffices to realise that since
$n=n(m,\theta)$ then at least locally one can express $m=m(n,\theta)$.
Performing a gradient expansion (i.e., in the expansion in $\kappa$)
one then has $\nabla_{\mu}m=\frac{1}{n_{m}}\nabla_{\mu}n+O(\kappa)$
with $n$ and $n_{m}$ taken at $\theta=0$, that is for the perfect
fluid limit. With this, the equation of motion (or charge conservation)
\eqref{eq:eomDiffusionDiv} reduces to \[
\dot{n}+\theta n=\nabla_{\lambda}\left(\frac{\kappa}{n_{m}m}\perp^{\mu\nu}\nabla_{\nu}n\right)\,+O(\kappa^{2}).\]
Bar the terms involving the expansion, this equation is the standard
diffusion equation \begin{equation}
\dot{n}=-\boldsymbol{\overline{\nabla}}_{\mu}\left(\mathfrak{D}\boldsymbol{\overline{\nabla}}^{\mu}n\right)+\mathfrak{D}a^{\mu}\boldsymbol{\overline{\nabla}}_{\mu}n\,,\label{eq:DiffEq}\end{equation}
with the diffusion coefficient a function of the diffusivity%
\footnote{It is useful to compare this expression for $\mathfrak{D}$ with \cite[Eq. (59.9), p. 232]{Landafshitz_Fluid}.
Note that our $\kappa/m=-\alpha$ and $\rho c=n$. %
} \begin{equation}
\mathfrak{D}=-\frac{\kappa}{n_{m}m}=-\frac{\kappa}{n}c_{\text{s}}^{2}\equiv -\tau_\mathfrak{D} c_\text{s}^2\,,\label{eq:DiffusionProper}\end{equation}
with $\tau_\mathfrak{D}$ the diffusion timescale appearing here since the diffusion coefficient has dimensions of L$^2$/T. In the limit of an incompressible fluid, Eq. \eqref{eq:DiffEq} is in fact the exact equation of motion to all orders in $\kappa.$ The
last term in this diffusion equation is proportional to acceleration
and manifests that the equation is written in a non-inertial accelerated
frame. 

 Note that in this discussion we have not used the Einstein equations. The whole procedure is based on the presence of a higher derivative in the equation of motion, and the corresponding $\nabla_{\mu}a^{\mu}$ in the equation of motion i.e.\ charge conservation \eqref{eq:eomDiffusionDiv}.

One may wonder whether the system gives enough freedom to ever realise this diffusive behaviour. It can be seen that this can occur when the initial data are such that the expansion $\theta$ vanishes and $\dot\theta$ is tuned to be small. This means that the timescale for the expansion of the volume is much larger than the timescale for  the diffusion, 
\begin{equation}
\tau_\mathfrak{D}\ll \theta^{-1}\,.
\end{equation}
This condition also happens to imply that the contribution from the expansion to the particle number is negligible. Then we must require that, in the charge-conservation equation \eqref{eq:eomDiffusionDiv}, the expansion term is negligible compared to divergence term, or roughly
\begin{equation}
	\theta n \ll \kappa \nabla_\mu a^\mu \qquad\Rightarrow \qquad \nabla_\mu a^\mu \gg \tau_\mathfrak{D}^{-2}\,.
\end{equation}
Finally, one needs to make sure that the configuration is maintained for sufficient time, i.e.\ that $\theta$ evolves slowly enough. Through the Raychaudhuri equation \eqref{eq:Raychaudhuri} we require that
\begin{align}
	\frac{\kappa}{n}\left|\nabla_\mu a^\mu\right| \gg \dot\theta \tau_\mathfrak{D} &\approx \left|\nabla_\mu a^\mu\right| \tau_\mathfrak{D} -\sigma_{\alpha\beta}\sigma^{\alpha\beta} \tau_\mathfrak{D}\notag\\
	\Rightarrow \nabla_\mu a^\mu &\gg \left|\nabla_\mu a^\mu -\sigma_{\alpha\beta}\sigma^{\alpha\beta}\right|\,,
\label{FineTuning}
\end{align}
where all the above quantities depend purely on initial data. The final condition is precisely the tuning that we are required to perform on these initial conditions to ensure that diffusion occurs for some finite time much longer than the characteristic diffusion timescale $\tau_\mathfrak{D}$.

This is a rather fine-tuned situation, which is already clear from the fact that this cancellation is not possible in 1+1 dimensions. However, initial conditions obeying \eqref{FineTuning} definitely exist. For example, in 1+3 dimensional Minkowski spacetime, these initial conditions can be easily realised
by the following ansatz in Cartesian coordinates $(t,\mathbf{x})$: $\partial_t\phi_0 (\mathbf{x})=1$ and $\phi_0(\mathbf{x}) = \epsilon f(\mathbf{x})$, where 
$f(\mathbf{x})$ is a harmonic function, $\Delta f(\mathbf{x}) =0$, and $\epsilon$ is a small parameter. In that case the geometric invariants are $\theta=\mathcal O(\epsilon^3)$, $ \nabla_\mu a^\mu=\mathcal O(\epsilon^2)$ and $\sigma^{\mu\nu}\sigma_{\mu\nu}=\mathcal O(\epsilon^2)$, whereas from the Raychaudhuri equation we get $\dot \theta = \mathcal O(\epsilon^4)$. 
\\
 We conclude that the interpretation of this system as a diffusive
one seems appropriate and robust. We emphasise that this is a non-dissipative
form of diffusion present at zero temperature.  Further one can see
that $\mathfrak{D}a^{\mu}\boldsymbol{\overline{\nabla}}_{\mu}n=-\mathfrak{D}\left(\boldsymbol{\overline{\nabla}}n\right)^{2}/\left(mn_{m}\right)$
where $\left(\boldsymbol{\overline{\nabla}}n\right)^{2}=-\perp^{\mu\nu}\boldsymbol{\overline{\nabla}}_{\mu}n\boldsymbol{\overline{\nabla}}_{\nu}n$.
So that Eq. \eqref{eq:DiffEq} is in fact a nonlinear diffusion
equation. 

Let us now turn to the identification of the imperfect part of the
EMT, which can be done by comparing with the conventional gradient
expansion performed in relativistic thermo/hydrodynamics. As we have
already mentioned in our discussion of the EMT in section \ref{sub:EMT},
it is necessary to solve the equation of motion for the scalar field
in order to express the EMT purely through variables on the Cauchy
hypersurface. In doing so, we rewrite the EMT in the effective form
of a gradient expansion with $\kappa$ playing the role of the expansion
parameter. Substituting the equation of motion \eqref{FluidEoM} and
the expression for the particle current in fluid variables \eqref{eq:nAsPm}
into the expression for the pressure \eqref{TotalPressure} gives
an EMT of the form\[
T_{\mu\nu}=\mathcal{E}u_{\mu}u_{\nu}-\left(\widetilde{P}-\zeta\theta\right)\perp_{\mu\nu}+u_{\mu}q_{\nu}+u_{\nu}q_{\mu}+\mathcal{O}(\kappa^{2})\,,\]
where we now have a pressure term corrected for the first-oder terms
in this expansion,\[
\widetilde{P}\equiv P+\frac{\kappa}{D}\mathcal{E}_{\phi}\,,\]
and there appears a bulk viscosity-like term with a would-be viscosity
coefficient \[
\zeta\equiv-D^{-1}\kappa P_{m}+\mathcal{O}(\kappa^{2})=-\kappa\left(\frac{n}{n_{m}}\right)+\mathcal{O}(\kappa^{2})\,,\]
and we have neglected the terms second-order in the diffusivity $\kappa$.
We can compare this expression with the sound speed that we have obtained
in \cite[Eq. 3.20]{Deffayet:2010qz} for the cosmological background.%
\footnote{As we show in our forthcoming work Ref.\ \cite{Metrics}, the sound
speed on general backgrounds is more complicated but gives the same
results at the level of precision required in this discussion.%
} Again dropping all terms of order $\kappa^{2}$, we can conclude
that when the correction from kinetic braiding is small, the effective
bulk viscosity, \begin{equation}
\zeta=-\kappa mc_{\text{s }}^{2}+\mathcal{O}(\kappa^{2})\,.\label{Viscosity}\end{equation}
However, it is important to bear in mind that this type of bulk viscosity
differs from the usual notion defined in terms of the conventional
gradient expansion. Indeed, in this expansion an important assumption
is that the energy density does not receive any corrections involving
gradients of potentials or velocities. In our fluid, however, we do
have a contribution to $\mathcal{E}$ proportional to $\theta$. Hence,
the only way to appropriately compare to the language of the gradient
expansion is to perform a redefinition of what is meant by the hydrodynamic
potentials, order by order in $\kappa$. Thus, one has to introduce
a new chemical potential
\begin{equation}
\hat{m}=\hat{m}(m,\theta)\,,
\label{newChemical}
\end{equation}
in such a way as to keep the form that is the energy density unchanged
between the zeroth and first orders in the expansion,
\begin{equation}
\mathcal{E}_{0}(\hat{m})\equiv\mathcal{E}(m,\theta)=\mathcal{E}_{0}(m)+\kappa m\theta\,,
\label{newEnergy}
\end{equation}
where $\mathcal{E}_{0}=mP_{m}-P$ is the k-\emph{essence} energy density.
All quantities are supposed to be expressed now in terms of $\hat{m}$.
At leading order, we therefore have\[
\hat{m}\approx m+\frac{\kappa m\theta}{\mathcal{E}_{0,m}}=m+\frac{\kappa\theta}{D}\,.\]
the charge density $n$ becomes a function of $\hat{m}$ only, up
to first order in $\kappa$
\begin{equation}
n(m)=n_{0}(m)+\kappa\theta=n_{0}(\hat{m})+\mathcal{O}(\kappa^{2})\,.
\label{newParticleNumber}
\end{equation}
This allows us to identify this modified chemical potential with the
definition\[
\hat{m}=\frac{\partial\mathcal{E}}{\partial n}\,,\]
which is valid to the same order up to same order without the need
to specify that the derivative is taken at constant volume. The pressure
using the new potential becomes
\begin{equation}
\widetilde{P}(\hat{m})=\widetilde{P}(m)+\kappa m c_{\text{s}}^{2}\theta\,.
\label{newPressure}
\end{equation}
When combined with the bulk viscosity term we obtained in Eq. \eqref{Viscosity},
we can see that the contributions proportional to the expansion $\theta$
cancel out, and therefore the correct identification for the bulk
viscosity of this fluid is \begin{equation}
\zeta_{\text{grad exp}}=0.\label{eq:ZeroZeta}\end{equation}
This is in tune with the fact that no actual dissipation takes place,
as indeed it must not in the case of a theory arising from an action
principle. \footnote{The same redefinitions of the chemical potential \eqref{newChemical}, particle number \eqref{newParticleNumber}, energy density \eqref{newEnergy} and pressure \eqref{newPressure} were recently rederived in the work \cite{Dubovsky:2011sj}, c.f.\ equations (A9)-(A12).}

Having said that, we must emphasise that the EMT does depend on the
expansion $\theta$: even the energy density does (in the original
variables). This corresponds to a kind of ``viscous'' behaviour: a
(perhaps new) type of nontrivial response to expansion that is not
dissipative. Hence, it still seems appropriate to call this a viscid
fluid, even if the conventionally defined linear bulk and shear viscosity
coefficients vanish. 

Taking as a measure of this type of viscosity the coefficient defined
above in Eq. \eqref{Viscosity}, we can see that there is also a relation
between viscosity and diffusion in this fluid: \[
\zeta=\mathfrak{D}nm\,.\]

\section{Equilibrium and Vacuum Configurations}%
\label{sec:Equilibrium}

In this section, we will discuss the conditions on the fluid that
are implied by the requirement that it is in equilibrium. As we will show
here, the coupling of the braided fluid to gravity imposes an additional
condition on the braided fluid compared to k-\emph{essence}.

Let us assume a non-vanishing number of particles in the fluid, $n\neq0$.
In equilibrium in a stationary gravitational field with a timelike
Killing vector $\xi_{\mu}$, the quantity which is constant throughout
a system (in our case---on the spacelike hypersurfaces $\Sigma_{\phi}$)
is not the locally measured chemical potential $m$ but \begin{equation}
m\sqrt{g_{\mu\nu}\xi^{\mu}\xi^{\nu}}=\mbox{const}\,.\label{ChemicalPotentialinEquilibrium}\end{equation}
This is sometimes called Tolman-Klein condition, see Ref. \cite{Klein}
and Ref. \cite[\S27, p. 77]{LandavshitzV}. 

In the discussion here we will concentrate on the shift-symmetric
case. Firstly, since $\phi$ is an \emph{intrinsic clock} and in
equilibrium no physical observables should depend on this \emph{intrinsic
clock} explicitly. Secondly, comoving observables like $\mathcal{E}$,
$m$, etc. should be independent of comoving time, e.g. $\dot{\mathcal{E}}=u^{\lambda}\nabla_{\lambda}\mathcal{E}=0$.
Since we are building our description on the assumption that $m=\dot{\phi}\neq0$,
it better be the case that $\phi$ is not an observable, at least
in equilibrium. 

The requirement that observables be independent of comoving time implies
that the Killing vector $\xi_{\mu}\propto u_{\mu}$. Since $u^\mu$ is hypersurface orthogonal by definition, this implies in fact that the equilibrium configuration is not just stationary, but also static. To satisfy Eq.
\eqref{ChemicalPotentialinEquilibrium} in an arbitrary curved spacetime
automatically---without new conditions---we have to assume that the
coefficient of proportionality is $m^{-1}$ so that \begin{equation}
\text{in equilibrium }\quad\xi_{\mu}=\frac{u_{\mu}}{m}\quad\text{is a Killing vector}.\label{EqulibriumKilling}\end{equation}
Therefore \[
\nabla_{\nu}\xi_{\mu}+\nabla_{\mu}\xi_{\nu}=\frac{2\mathcal{K}_{\mu\nu}}{m}-\frac{2u_{\mu}u_{\nu}}{m^{2}}\dot{m}=0\,,\]
a projection of which gives \begin{eqnarray}
\dot{m}=0 & \mbox{and} & \mathcal{K}_{\mu\nu}=0\,.\label{EquilibriumConditions}\end{eqnarray}
In particular, the expansion must vanish, $\theta=\mathcal{K}_{\mu}^{\mu}=0$.
Thus we have shown that, in equilibrium, the motion of the fluid should
be rigid in the Born sense: $\mathcal{K}_{\mu\nu}=0$. This is the
standard requirement, see e.g. Ref. \cite{Israel:1976tn}.

We can also consider situations where more than one type of matter
is present. Since $\xi^{\mu}$ is the only time-like Killing vector,
the velocity of the external fluid must lie along $\xi^{\mu}$, and
since the velocities are normalised---$u_{\text{ext}}^{\mu}=u^{\mu}$.
The Tolman-Klein condition Eq. \eqref{ChemicalPotentialinEquilibrium}
then implies chemical potentials of all the constituents must be equal,
$m_{\text{ext}}=m$. 

As we have already mentioned, the chemical potential $m$ for the
system in an external field can be space dependent even in equilibrium.
Thus the acceleration does not necessarily vanish. However, this is
a result of the fact that a non-zero acceleration implies that $u^{\mu}$
is not tangent to a geodesic: the energy flux in equilibrium appears
since the LRF is non-inertial. One should stress, however, that the
time dependence of the acceleration does vanish. The appropriate measure
of the time dependence of a vector in the Lie derivative, and in this,
case
 \[
\pounds_{\xi}a_{\mu}=a^{\lambda}\nabla_{\mu}\left(\frac{u^{\lambda}}{m}\right)+\frac{u^{\lambda}}{m}\nabla_{\lambda}a_{\mu}=0\,.\]

In general, to find the equilibrium configuration as a function on
space, one would need to apply the equilibrium conditions Eq. \eqref{EquilibriumConditions}
to the Raychaudhuri equations and Einstein equations and solve this
system. This is potentially a rich problem and lies outside of the
scope of this work. Nonetheless, let us elucidate one significant
difference between k-\emph{essence} and the braided fluid. Under the
equilibrium conditions, the equation of motion for k-\emph{essence}
is trivial and does not supply any new information. However, in the
case of the braided fluid, the equation of motion \eqref{eq:EoM}
reduces to a new constraint \begin{equation}
\kappa R_{\mu\nu}u^{\mu}u^{\nu}=-\kappa_{m}ma^{2}\,,\label{k_vs_R}\end{equation}
which, using the Einstein equations, becomes
\begin{equation}
\mathcal{E}+3P+\mathcal{E}_\text{ext}+3P_\text{ext}=-2(\ln\kappa)'a^{2}\,,
\label{eq:equil}
\end{equation}
where the prime denotes differentiation with respect to $\ln m$ and the quantities with the subscript external are any potential contributions to the EMT coming from matter external to the braided scalar.
Let us just say that this constraint imposes a condition on the equation
of state of the braided scalar in equilibrium. In the extreme case
of a constant diffusivity $\kappa$, this constraint implies that
the equation of state is such that the scalar arranges itself in a
configuration which screens the gravitational effect of the mass of
the total EMT. In particular, when there is no external matter, the
equation of state in equilibrium is $w=-1/3$. This is another facet
of the monitoring behaviour present in the cosmological solutions
which we discussed in detail in Ref. \cite{Deffayet:2010qz}.

\medskip{}
Let us now briefly mention vacuum configurations where the charge
density vanishes, $n=0$. In Ref. \cite{Deffayet:2010qz} we showed
that for the shift-symmetric theories with \emph{kinetic gravity braiding}
there exist such non-trivial vacuum solutions and that they are dynamical
attractors in an expanding universe. As one can see from Eq. \eqref{eq:particle density},
the vanishing of the shift-charge implies that the motion is restricted
to obey \[
\theta\left(m\right)=-\frac{P_{m}}{\kappa}\,.\]
The structure of these configurations is potentially even richer than that of equilibrium
configurations and this issue deserves a further detailed analysis
going far beyond of the scope of this paper.

\section{How unnaturally complicated could it be?}%
\label{sec:How-unnaturally-complicated}

One really interesting example of how unexpectedly complicated and
unnatural the function $K\left(X\right)$ can be is provided by the
noninteracting relativistic degenerate fermions with mass $m_{\text{e}}$
and chemical potential $m$, see e.g. \cite[\S61, p. 180]{LandavshitzV}.
In equilibrium, we have
\begin{align}
 & n=\frac{m_{\text{e}}^{3}}{3\pi^{2}}\sinh^{3}\frac{\xi}{4}\,,\\
 & P=\frac{m_{\text{e}}^{4}}{32\pi^{2}}\left(\frac{1}{3}\sinh\xi-\frac{8}{3}\sinh\frac{\xi}{2}+\xi\right)\,,\notag\\
 & \mathcal{E}=\frac{m_{\text{e}}^{4}}{32\pi^{2}}\left(\sinh\xi-\xi\right)\,,\notag\\
 & m=m_{\text{e}}\cosh\frac{\xi}{4}\,,\notag\end{align}
where $\xi$ is related to Fermi momentum $p_{F}$ as \[
\xi=4\text{arcsinh}\frac{p_{F}}{m_{\text{e}}}\,.\]
First of all, one can check that the Euler relation, Eq. \eqref{eq:Euler}, indeed holds. Further,
we can re-express the pressure through the chemical potential $m$
so that we obtain\[
P\left(m\right)=\frac{m_{\text{e}}^{4}}{12\pi^{2}}\left[\frac{3}{2}\ln\left[\frac{m}{m_{\text{e}}}+\sqrt{\left(\frac{m}{m_{\text{e}}}\right)^{2}-1}\right]+\sqrt{\left(\frac{m}{m_{\text{e}}}\right)^{4}-\left(\frac{m}{m_{\text{e}}}\right)^{2}}\left[\left(\frac{m}{m_{\text{e}}}\right)^{2}-\frac{5}{2}\right]\right]\,.\]
If we restrict our attention to the simplest motion---vorticity-free---
we have to follow Schutz \cite{Schutz} and the identification used in this work, Eq.  \eqref{eq:effectiveM},
and substitute $m=\sqrt{2X}=\sqrt{\left(\partial\phi\right)^{2}}$
into this formula for the pressure to obtain the Lagrangian. As the
result of this procedure, we get a rather unusual scalar field
theory of the k-\textit{essence} type:\[
P\left(\partial\phi\right)=\frac{m_{\text{e}}^{4}}{12\pi^{2}}\left[\frac{3}{2}\ln\left[\sqrt{\frac{\left(\partial\phi\right)^{2}}{m_{\text{e}}^{2}}}+\sqrt{\frac{\left(\partial\phi\right)^{2}}{m_{\text{e}}^{2}}-1}\right]+\sqrt{\frac{\left(\partial\phi\right)^{4}}{m_{\text{e}}^{4}}-\frac{\left(\partial\phi\right)^{2}}{m_{\text{e}}^{2}}}\left[\frac{\left(\partial\phi\right)^{2}}{m_{\text{e}}^{2}}-\frac{5}{2}\right]\right]\,.\]
Note that this Lagrangian describes a simple physical system and the incorporation
of any interaction would only serve to make the structure more complicated.
Moreover, this Lagrangian does not have a well defined vacuum limit
at $\left(\partial\phi\right)^{2}\rightarrow0$. This is the manifestation
of the simple physical fact that the fermion chemical potential can
never be smaller than the mass of the fermion $m_{\text{e}}$. Even in the limit $\left(\partial\phi\right)^{2}\rightarrow m_{\text{e}}^{2}$
the asymptotic is not analytic: \[
P\left(\partial\phi\right)\simeq\frac{4\sqrt{2}}{15\pi^{2}}m_{\text{e}}^{4}\left(\frac{\sqrt{\left(\partial\phi\right)^{2}}}{m_{\text{e}}}-1\right)^{5/2}\,.\label{eq:fermscalar}\]
Observe that the fundamental degrees of freedom in
this system are fermions which have nothing to do with the effective
bosonic field $\phi$. Moreover, the Lagrangian does not appear to
be technically natural and the only free parameter $m_{\text{e}}$
does not correspond to the actual strong-coupling scale. One could
expect that the latter is governed by the scale of particle separation, $n^{1/3}$, which can be exponentially larger than $m_{\text{e}}$. 

It is important to stress here that the Lagrangian Eq.~\eqref{eq:fermscalar} does not model all the degrees of freedom which exist in an ideal Fermi gas. In particular, it does not contain the zero sound, which is a non-equilibrium propagating mode.\footnote{We thank Gregory Gabadadze, Massimo Porrati and Alberto Nicolis for discussing this with us.} In ideal Fermi gases this mode dominates at low temperature, while the adiabatic hydrodynamic mode modeled by Eq.~\eqref{eq:fermscalar} is actually suppressed. However, the addition of even a small attractive interaction allows the adiabatic mode to propagate \cite[p.~163]{LandavshitzIX}. Adding such a small attractive interaction should have only a small effect on the equation of state of the fluid, since it only affects those modes lying close to the Fermi surface. On the other hand, we do not know whether there exist real examples of fluids where the hydrodynamic description we are using above would dominate over the non-equilibrium zero-sound modes. However, e.g. the modeling of ultra-high-density neutron stars is usually performed this hydrodynamic approximation.
 
Having seen the level of complexity allowed even for the simplest
systems one can ask what are the natural structures of the function
$G\left(X\right)$? The simplest option for the diffusivity $\kappa$
is just to be constant. This corresponds to \begin{eqnarray*}
G\left(X\right)=\frac{\kappa}{2}\ln X\,, & \text{for} & \kappa=\text{const}\,.\end{eqnarray*}
Another option would be to declare the diffusion coefficient $\mathfrak{D}$
to be constant. This would correspond to $G$'s and $K$'s being connected
as\begin{eqnarray*}
G\left(X\right)=-\mathfrak{D}\sqrt{2X}K_{X}\,, & \text{for} & \mathfrak{D}=\text{const}\,,\end{eqnarray*}
where we have omitted a constant of integration. In this case, $G$
is just proportional to the charge density in equilibrium. However,
as the simple physical example with fermions teaches us, Nature
may not enjoy such structural simplicity at all.

\section{Discussion and Future Directions}%
\label{sec:Conclusions}

In this paper, we used a hydrodynamical language to describe the dynamics
of theories with \emph{kinetic gravity braiding}. This fluid picture
turns out to be extremely useful: it provides an intuitively clear
physical meaning to otherwise obscure combinations of derivatives
of the field and the Lagrangian. This drastically simplifies notation
allowing a much better understanding of the system. As is well known,
finding the correct variables in many cases provides an easy path
to the solution. Moreover, this notation clarifies the information
encoded in the free functions $K\left(\phi,X\right)$ and $G\left(\phi,X\right)$,
making it possible to construct these theories based on physical principles
and not solely from a naive naturalness. 

The key feature of our fluid, is the dependence of the charge density
and, correspondingly, the energy density on the expansion $\theta$.
This allows for the presence of bulk viscosity-like effects and diffusion
without any dissipation. This novel property, to our best knowledge,
had not been considered in the literature of hydrodynamics hitherto. 

We would like to mention that our identification of shift charges
with particles and, therefore, of the proper-time gradient of the scalar field with the chemical potential is not actually unique. In fact, we could also have assumed that these
charges correspond to entropy. From the first law of thermodynamics,
it follows that this alternative identification could be obtained by the exchange $n\rightarrow s$
and $m\rightarrow T$, where $T$ is the temperature. We find that
the identification presented in the paper is more satisfactory. In the case with entropy, even for a perfect fluid EMT, a non-shift symmetric Lagrangian would imply that entropy is not conserved. Moreover, even in the shift-symmetric case, our diffusion flux would correspond
to a heat flux, but one which would nonetheless conserve entropy.  

The equation of motion for the scalar is a highly nonlinear partial
differential equation of the Amp\`{e}re-Monge type. This class of
equations belongs to the frontiers of current research in mathematics.
In fact, contrary to the quasilinear case, it is not even clear under
which conditions the Cauchy problem is well posed. It may happen that
this fluid picture could aid in developing the understanding of this
problem. In the forthcoming paper, Ref. \cite{Metrics} we make a
first step in this direction by discussing the high-frequency stability
and acoustic geometry in these theories. 

As we have alluded to already, our hydrodynamical picture owes a lot
to Ref. \cite{Schutz}, where perfect fluids with vorticity and thermal
effects are described through a Lagrangian. It would be interesting
to investigate whether a non-pathological and meaningful generalisation
of perfect fluids with vorticity and thermal properties is possible
along the lines discussed in our paper. We believe that for
this purpose the Lagrangian $\tilde{\mathcal{P}}\left(\phi,m,\theta\right)$
from Eq. \eqref{eq:Action_1Order} could be rather suitable. In Ref.
\cite{Schutz} the author had already conjectured: {}``It may also
be possible to extend this work to viscous fluids...''. In fact, the
current paper can be considered as a first step in this direction.
Moreover, one could consider similar generalisations in the framework
of Ref. \cite{Dubovsky:2005xd}; for a recent development in this
direction see e.g. Refs \cite{Nicolis:2011ey,Endlich:2010hf}. 

In a rather natural extension, one could employ the formalism developed
in this paper to consider the hydrodynamical properties of the more
general \emph{galileons }and theories such as those of Ref. \cite{Deffayet:2011gz}.
There is, however, a potential difficulty in interpretation which
is related to a necessary appearance of curvature terms in the action,
see Ref. \cite{Deffayet:2009wt,Deffayet:2009mn}. 

Further, one should not forget that boundary terms are in fact unavoidable
for the proper formulation of the action principle in theories with
higher derivatives, see e.g. \cite{Dyer:2009yg,Dyer:2008hb}. It would
be very interesting and important to find these terms and their possible
fluid interpretation in theories with \emph{kinetic gravity braiding. }

Having in mind the higher-derivative structure of the theory, it is
interesting to speculate how one could change the fluid picture using
constraints as in Ref. \cite{Lim:2010yk}. It seems that, in particular,
one could obtain a \emph{kinetically braided dust}---a braided fluid
moving along geodesics. Another option would be to constrain the expansion
in the action and obtain in that way a braided incompressible fluid
with diffusion. 

Finally, the fluid presented here appears to naturally act so as to screen
the gravitational effects of matter when it is in an equilibrium configuration.
This is another manifestation of the monitoring effect that we discussed
for the cosmological solutions of these models in Ref. \cite{Deffayet:2010qz}.
Such phenomenology would clearly be important in cosmology during
non-linear structure formation. Could the equilibrium configurations
which we have described here provide a model for dark-matter haloes
which significantly deviates from the standard CDM paradigm? 

To conclude, we are in these days witnessing a revival of interest
and a rather exciting series of developments in our understanding
of non-canonical field theories and hydrodynamics. The bestiary of
models used in cosmological models has grown and we are in the process
of trying to understand the implications. Eventually, these models
will need to stand up to observational tests, whether in cosmology
or possibly even in some condensed-matter systems, and we are in the
process of trying to understand whether they could have any completely
new signatures for which to hunt. 

\acknowledgments{\label{sec:Thanks}
\addcontentsline{toc}{section}{Acknowledgments}
It is a pleasure to thank Luca Amendola, Eugeny Babichev, Andrei Barvinski,
Ram Brustein, Gerhard Buchalla, Jorge Casalderrey-Solana, Alexander Dolgov, Sergei Du\-bov\-sky, Gia Dvali, Andrei Frolov, Gregory Gabadadze, Jaume Garriga,
Cristiano Germani, Dmitry Gorbunov, Andrei Gruzinov, Bhilahari Jeevanesan,
Andrei Khmelnitsky, Maxim Libanov, Viatcheslav Mukhanov, Shinji Mukohyama,
Alberto Nicolis, Massimo Porrati, Valery Rubakov, Mikhail Shaposhnikov, Sergey Sibiryakov, Christof
Wetterich for very useful discussions and criticisms. We extend special
gratitude to C\'edric Deffayet for his comments throughout the preparation
of this work and for the collaboration on initial stages of this project.
A significant part of this work was completed at the time when I.~S.
and A.~V. were enjoying the support of the James Arthur Fellowship
at the Center for Cosmology and Particle Physics, New York University.
I.~S. and A.~V. would like to thank for this support and for the
lively atmosphere at the Center. In the final stages of this project,
I.~S. was supported by the DFG through TRR33 ``The Dark Universe''.
I.~S. and A.~V. would also like to thank the CERN Theory Division
for their hospitality during the preparation of this manuscript. This
visit of A.~V. to CERN was supported through a grant of the David
and Lucile Packard Foundation. A.~V. is also thankful to the organizers
of the workshop \emph{Gravity and Cosmology 2010} at the Yukawa Institute
for Theoretical Physics, Kyoto University for their hospitality and
financial support during the intermediate stages of this project,
many results of which were presented and discussed during the workshop. 
}
\appendix

\section{Choice of Frames}%
\label{sec:frames}
 
The choice of $u^{\mu}$ defined in Eq. \eqref{eq:4velocity} as a
velocity for the frame is special in that it is explicitly vorticity
free as a result of this velocity's being parallel to the gradient
of $\phi$. This means that we are permitted to perform a foliation
of the space-time using $\phi$ as an internal clock and the surfaces $\Sigma_{\phi}$:
$\phi=\mathrm{const}$ as our spatial hypersurfaces. This particular
foliation is then a natural candidate for being a Cauchy surface.
In the follow-up work Ref. \cite{Metrics}, we discuss the requirements
on the possible general fluid configurations which ensure that this
particular choice of spatial hypersurfaces indeed provides a surface
on which initial values can be supplied in the usual unconstrained
fashion and one can honestly describe the scalar field as a fluid. 

Using a general frame moving with some other velocity $U^{\mu}$ we
could have defined the pressure, energy density and energy flow using
equations \eqref{EnergyDensity}, \eqref{EquilibriumPressure} and
\eqref{eq:EnergyFlow} with the substitution $u^{\mu}\rightarrow U^{\mu}$.
In these variables the EMT would be \[
T_{\mu\nu}=\epsilon_{U}U_{\mu}U_{\nu}-\perp_{\mu\nu}^{U}\mathcal{P}_{U}+q_{\mu}^{U}U_{\nu}+q_{\nu}^{U}U_{\mu}+\pi_{\mu\nu}\,,\]
 where \[
\pi_{\mu\nu}=\left(\perp_{\mu\alpha}^{U}\perp_{\nu\beta}^{U}-\frac{1}{3}\perp_{\mu\nu}^{U}\perp_{\alpha\beta}^{U}\right)T^{\alpha\beta}\,,\]
is the shear-stress tensor. Note that in principle we could perform
this decomposition for an arbitrary EMT. Generically, none of the
shear, rotation or vorticity will vanish in such a frame and the shear-stress
tensor will contain terms with \emph{$\mathcal{P}$} and, correspondingly,
$\dot{m}$. Hence, for general initial data, $\pi_{\mu\nu}$ will
not vanish. In particular, this is the case for the two most popular
frames used in the analysis of standard theories of imperfect fluids:
the Eckart frame, where the frame moves with the particle flow (so
that $U_{\text{E}}^{\mu}=J^{\mu}/\sqrt{J^{\lambda}J_{\lambda}}$),
and the Landau-Lifshitz frame, which moves together with the energy
so that $T_{\nu}^{\mu}U_{\text{LL}}^{\nu}=\epsilon_{\text{LL}}U_{\text{LL}}^{\mu}$.
It could well be that these frames may not even exist for some otherwise
reasonable initial data. 

One should note, however, that up to (and
including) the first order in $\kappa$, these two frames coincide, since 
the actual heat flux is absent, see \cite[Eq. (2), p. 312]{Israel:1976tn}. Moreover, up to this order, the energy-momentum tensor in this Landau-Lifshitz frame takes the perfect-fluid form. And indeed, up to and including the first order in $\kappa$ we are then dealing with a perfect fluid. However, once higher-order corrections are included, we see that the price paid for this frame change is that the fluid flow is now no longer vorticity free, with corrections also coming at second order in $\kappa$. We show this explicitly below. 

Let us for simplicity concentrate on the shift-symmetric case. In the Eckart frame (suppose it exists so that \eqref{EckartExists} holds), the shift-charge density is 
\[
n_{\text{E}}=\sqrt{J_{\mu}J^{\mu}}=n+\mathcal{O}\left(\kappa^{2}\right)=P_{m}+\kappa\theta+\mathcal{O}\left(\kappa^{2}\right)\,.
\]
The 4-velocity of this frame is 
\[
U_{\text{E}}^{\mu}=J^{\mu}/n_{\text{E}}=u^{\mu}+\mathcal{O}\left(\kappa\right)\,.
\]
Now let us calculate the \textit{kinematical rotation vector} for
the Eckart frame, see e.g. \cite{Gourgoulhon:2006bn}:

\[
\omega^{\mu}\left(U_{\text{E}}\right)\equiv\varepsilon^{\alpha\beta\gamma\mu}\left(\nabla_{\alpha}U_{\text{E}}{}_{\beta}\right)U_{\text{E}\gamma}=\omega^{\mu}\left(J\right)/n_{\text{E}}^{2}\,.
\]
This kinematical rotation vector is related to the previously mentioned
rotation tensor or twist in this frame $w_{\mu\nu}^{\text{E}}=1/2\left(\bot_{\mu\alpha}^{U}\nabla^{\alpha}U_{\text{E}\nu}-\bot_{\nu\alpha}^{U}\nabla^{\alpha}U_{\text{E}\mu}\right)$
in the following way: 
\[
\omega^{\mu}\left(U_{\text{E}}\right)=\varepsilon^{\alpha\beta\gamma\mu}\left(\bot_{\alpha\lambda}^{U}\nabla^{\lambda}U_{\text{E}}{}_{\beta}\right)U_{\text{E}\gamma}=\varepsilon^{\alpha\beta\gamma\mu}w_{\alpha\beta}^{\text{E}}U_{\text{E}\gamma}\,.
\]
Further for $\omega^{\mu}\left(J\right)$ we have 
\[
\omega^{\mu}\left(J\right)=\varepsilon^{\alpha\beta\gamma\mu}\nabla_{\alpha}J_{\beta}J_{\gamma}\,.
\]
Some tedious but straightforward algebra yields the result that the vorticity in the Eckhart frame, in terms of the variables in our chosen local rest frame defined by the gradient of $\phi$,
\begin{equation}
	\omega^\mu(U_\text{E}) = \frac{\kappa^2}{n_\text{E}^2} \epsilon^{\alpha\beta\gamma\mu} S_\alpha u_\beta a_\gamma\,,
\end{equation}
with
\begin{equation}
	S_\alpha \equiv \frac{\boldsymbol{\overline{\nabla}}_{\alpha}\dot{m}}{m} - \boldsymbol{\overline{\nabla}}_{\alpha}\theta\,.
\end{equation}
Thus the kinematical rotation vector $\omega^{\mu}\left(U_{\text{E}}\right)$
is $\mathcal{O}\left(\kappa^{2}\right)$ and vanishes on equilibrium configurations. In order to calculate it, one must solve the equation of motion \eqref{FluidEoM}. Nonetheless, for general initial
data,  neither $\boldsymbol{\overline{\nabla}}_{\alpha}\dot{m}$ nor $\boldsymbol{\overline{\nabla}}_{\alpha}\theta$ are parallel
to $a_{\mu}$, therefore $\omega^{\mu}\left(U_{\text{E}}\right)$
does not vanish. Nonvanishing vorticity implies (see Frobenius theorem) that the Eckart frame is not suitable for the formulation of the Cauchy problem, because $U_{\text{E}}^{\mu}$
is not hypersurface-orthogonal. 

Using the variables we defined in section~\ref{sec:Diffusion}, we can also rewrite the full EMT
\begin{equation}
	T_{\mu\nu} = (\mathcal{E}_0+\widetilde{P})u_\mu u_\nu -g_{\mu\nu}\widetilde{P}-\kappa \hat{m} (u_\mu a_\nu + u_\nu a_\mu) + \mathcal{O}(\kappa^2)\,. \label{eq:EMT_Ok}
\end{equation}
We can now perform a diagonalization into the Landau-Lifshitz frame by boosting the velocity $u_\mu$ along $a_\mu$,
\begin{align}
	U_{\text{LL}\mu} &=  u_\mu  - \alpha a_\mu + \mathcal{O}(\kappa^2) \\
	\widetilde{a}_\mu &= -\alpha u_\mu  +a_\mu +\mathcal{O}(\kappa^2)\notag \,,
\end{align}
with $\alpha \equiv \kappa \hat{m}/(\mathcal{E}_0+\widetilde{P})$. Rewriting Eq.~\eqref{eq:EMT_Ok} using the new velocity, we obtain an EMT of perfect-fluid form,
\begin{equation}
	T_{\mu\nu} = (\mathcal{E}_0+\widetilde{P})U_{\text{LL}\mu} U_{\text{LL}\nu} - g_{\mu\nu}\widetilde{P} + \mathcal{O}(\kappa^2)\,.
\end{equation}
However, at second order in $\kappa$, deviations away from a perfect fluid reappear in this frame in the form of a non-vanishing anisotropic stress proportional to $\widetilde{a}_\mu\widetilde{a}_\nu$.

The formulation of the theory in terms of the field $\phi$ provides
a natural reference frame which is neither the Landau-Lifshitz frame
nor the Eckart frame: the bulk velocity $u^{\mu}$ is neither the
velocity of energy nor the velocity of particles. Contrary to the
usual imperfect fluids, in this frame, the energy density and particle
density contain first derivatives of this four-velocity. Moreover,
the pressure contains the second time derivative of the scalar field.
It is also in this reference frame where the theory has a cosmological
solution and reduces to perfect-fluid configurations. Finally, the
shift-charge/particle-number current depends on initial data only
and it seems that there are no general restrictions on the functions
$G\left(\phi,X\right)$ and $K\left(\phi,X\right)$ such that the
current $J^{\mu}$ is timelike for all admissible initial data.

\section{Action Without Second Time Derivatives}%
\label{sec:FirstOrderAction}

Let us start from classical mechanics and consider the following
one-dimensional version of the system represented by our action \eqref{e:action},
\[
S=\int\mbox{d}t\left(K\left(q,\dot{q}\right)+G\left(q,\dot{q}\right)\ddot{q}\right)\,.\]
Now we can add a total derivative to this action without changing
the equation of motion so that: \[
\ddot{q}G\left(q,\dot{q}\right)+\frac{d}{dt}f\left(q,\dot{q}\right)=\ddot{q}\left(G+f_{,\dot{q}}\right)+f_{,q}\dot{q}\,.\]
We can eliminate the second derivative $\ddot{q}$ if we take \[
f=-\int^{\dot{q}}dvG\left(q,v\right)\,,\]
in which case the new, equivalent, Lagrange function without higher
derivatives is\[
L_{1}\left(q,\dot{q}\right)=K\left(q,\dot{q}\right)-\dot{q}\int^{\dot{q}}dvG_{,q}\left(q,v\right)\,.\]

Now let us see how one can eliminate the second time derivative for
our kinetically braided scalar field $\phi$. Let us use the Lagrangian
\eqref{e:L2} and write the action in the comoving frame in the form
\[
S_{\phi}=\int\mbox{d}^{3}x\int\mbox{d}\tau\sqrt{-g}\left(P-\kappa\dot{m}\right)\,,\]
where now $g=g^{\tau\tau}\perp$ with $\perp\equiv\mbox{det}\perp_{\mu\nu}$
and \[
g^{\tau\tau}=g^{\mu\nu}\partial_{\mu}\tau\partial_{\nu}\tau\,.\]
 Recalling the definition of the effective mass $m$ Eq. \eqref{eq:effectiveM}
we have\begin{eqnarray*}
\mbox{d}\tau=\frac{\mbox{d}\phi}{m}\,, & \text{so that} & g^{\tau\tau}=\frac{g^{\mu\nu}\partial_{\mu}\phi\partial_{\nu}\phi}{m^{2}}=1\,.\end{eqnarray*}
 Now we can add an arbitrary total derivative $\mbox{d}\left(f\sqrt{-\perp}\right)/\mbox{d}\tau$
without changing physics so that \[
S_{\phi1}=\int\mbox{d}^{3}x\int\mbox{d}\tau\sqrt{-\perp}\left(P-\kappa\dot{m}+f_{\phi}m+f\frac{\mathrm{d}}{\mathrm{d}\tau}\ln\sqrt{-\perp}+f_{m}\dot{m}\right)\,.\]
Therefore for \[
f\left(\phi,m\right)=\int^{m}dm'\kappa\left(m',\phi\right)\,,\]
we eliminate $\dot{m}=\ddot{\phi}$ from the action. Further, we notice
that the expansion can be expressed as \[
\theta=\frac{d}{d\tau}\ln\sqrt{-\perp}\,.\]
 Thus we obtain an equivalent new Lagrangian $\mathcal{\tilde{P}}$
without second time derivatives in comoving coordinates \begin{equation}
S_{\phi1}=\int\mbox{d}^{4}x\sqrt{-g}\left(P+mf_{\phi}+f\theta\right)=\int\mbox{d}^{4}x\sqrt{-g}\tilde{\mathcal{P}}\left(\phi,m,\theta\right)\,.\label{eq:Action_1Order}\end{equation}
Finally one can check that the difference between Lagrangians is indeed
a total derivative: \[
\mathcal{\tilde{P}}-\mathcal{P}=\kappa\dot{m}+m\int^{m}dm'\kappa_{\phi}+\theta\int^{m}dm'\kappa=\nabla_{\mu}\left(\int^{m}dm'\kappa\left(m',\phi\right)u^{\mu}\right)\,.\]
Of course if we were to change the coordinates away from this natural
LRF frame, $\mathcal{\tilde{P}}$ would again contain second time
derivatives.

\bibliographystyle{utphys}
\addcontentsline{toc}{section}{References}
\bibliography{KGB}

\end{document}